\documentclass[fleqn,usenatbib]{mnras}

\usepackage{newtxtext,newtxmath}
\usepackage[T1]{fontenc}
\usepackage{fix-cm}

\DeclareRobustCommand{\VAN}[3]{#2}
\let\VANthebibliography\thebibliography
\def\thebibliography{\DeclareRobustCommand{\VAN}[3]{##3}\VANthebibliography}

\usepackage{graphicx}	% Including figure files
\usepackage{amsmath}	% Advanced maths commands
\usepackage{subcaption}
\newcommand\mkms{\rm km \, s^{-1}}
\title[Environmental dependence of the MZR]{Environmental dependence of the Mass–Metallicity Star Formation Relations at $z=4-10$ with JWST}

\author[Li et al.]{
Qiong Li\thanks{ qiong.li@manchester.ac.uk}$^{1}$,
Christopher J. Conselice$^{1}$, 
Lewi Westcott$^{1}$,
Duncan Austin$^{1}$,
Tom Harvey$^{1}$,
\newauthor
Nathan Adams$^{1}$,
Vadim Rusakov$^{1}$,
James Arcidiacono$^{1}$,
Caio Moreira Goolsby$^{1}$
Chandana Hegde$^{2,3}$
\newauthor
Shuqi Fu$^{1,4,5}$
\\
$^{1}$ Jodrell Bank Centre for Astrophysics, University of Manchester, Oxford Road, Manchester UK \\
$^{2}$  DTU Space, Technical University of Denmark, Elektrovej 327, DK2800 Kgs. Lyngby, Denmark \\
$^{3}$ Cosmic Dawn Center (DAWN), Copenhagen, Denmark \\
$^{4}$ Department of Astronomy, School of Physics, Peking University, Beijing 100871, People's Republic of China \\
$^{5}$ Kavli Institute for Astronomy and Astrophysics, Peking University, Beijing 100871, People's Republic of China
}

\date{Accepted XXX. Received YYY; in original form ZZZ}

\pubyear{\the\year{}}

\begin{document}
\label{firstpage}
\pagerange{\pageref{firstpage}--\pageref{lastpage}}
\maketitle

% Abstract of the paper
\begin{abstract}We study how environment affects the mass--metallicity relation (MZR) at $z=4$--$10$ using deep imaging and spectroscopy from the \textit{James Webb Space Telescope} (\textit{JWST}). Combining CEERS and JADES, we compile a sample of 225 galaxies with stellar masses, star-formation rates, and gas-phase metallicities. We characterize environment using the projected fifth-nearest-neighbour surface density, $\Sigma_{5}$, within $\Delta z=\pm0.25$. At $4.5<z<7$, we find that galaxies in dense regions are more metal-rich at fixed $M_\star$ by $\sim0.1$--0.2\,dex, while the slopes of the MZR remain similar across environments. Including SFR increases the separation, suggesting more efficient chemical enrichment in overdense regions. Compared to the local $T_e$-based FMR, our full sample lies $\simeq0.2$--0.3\,dex below the $z\sim0$ relation, with a smaller deficit in overdense environments. We also examine how metallicity relates to galaxy size using NIRCam-based effective radii. Metallicity increases weakly with size up to $R_e\sim1$\,kpc and then flattens, with only a modest residual trend at fixed $M_\star$ and little environmental dependence. Using mass-weighted stellar ages at $5<z<10$, we find a positive age--metallicity relation in both environments, steeper in the field. Finally, we find that the star-formation rate density is higher in overdense regions at $z\simeq6$--9 by a factor of $\sim2$--3. Overall, our results suggest that environment accelerates both star formation and chemical enrichment during the epoch of reionization. Future wide-area JWST spectroscopy, combined with ALMA and Euclid, will better constrain the role of environment in early galaxy evolution.
\end{abstract}

% Select between one and six entries from the list of approved keywords.
% Don't make up new ones.

%%%%%%%%%%%%%%%%%%%%%%%%%%%%%%%%%%%%%%%%%%%%%%%%%%

%%%%%%%%%%%%%%%%% BODY OF PAPER %%%%%%%%%%%%%%%%%%

\section{Introduction}

The chemical enrichment of galaxies encodes the integrated history of star formation, gas inflow, outflow, and feedback, and therefore provides a fundamental probe of baryon cycling in galaxy formation. 
Among the empirical relations connecting these processes, the mass--metallicity relation (MZR) stands out as one of the most robust and ubiquitous correlations, linking stellar mass and gas-phase metallicity across a wide range of masses and cosmic epochs \citep{Tremonti2004,Maiolino2008,Curti2020,Sanders2021,Nakajima2023,Langeroodi2023,Stephenson2024,Pallottini2025}. 
In the local Universe, the MZR is well established and is commonly interpreted as the outcome of the interplay between metal production, gravitational potential depth, gas accretion, and feedback-driven outflows \citep[e.g.][]{Finlator2008,Lilly2013}. 
However, the physical origin of the MZR remains debated: while classical inflow–outflow models attribute the relation to mass-dependent metal retention and gas dilution \citep{Erb2006,Finlator2008,Spitoni2010,Bassini2024,PerezDiaz2024}, alternative interpretations emphasize the role of stellar mass as a tracer of total metal production \citep{BakerMaiolino2023}, secular evolution \citep{SomervilleDave2015}, stellar population age \citep{DuartePuertas2022}, or AGN feedback \citep{Thomas2019}.

At higher redshifts, the MZR evolves toward systematically lower metallicities at fixed stellar mass, reflecting the chemically immature nature of galaxies in the early Universe. 
Before the advent of the \textit{James Webb Space Telescope} (\textit{JWST}), observational constraints on the MZR were largely limited to $z\lesssim3$--4 due to the difficulty of accessing rest-frame optical emission lines \citep[e.g.][]{Maiolino2008,Sanders2018}. 
\textit{JWST} has fundamentally transformed this landscape by enabling direct metallicity measurements for statistically significant samples of galaxies at $z>4$ using rest-frame optical diagnostics \citep[e.g.][]{Nakajima2023,Curti2023,Curti2024,Fujimoto2024}. 
Recent studies consistently report a rapid decline in metallicity with redshift and substantial intrinsic scatter, indicative of highly dynamic, gas-rich systems undergoing rapid assembly and intense baryon cycling during the epoch of reionization.

The MZR is further extended by incorporating star formation rate (SFR) as a third parameter, leading to the so-called fundamental metallicity relation (FMR). 
In the local Universe, galaxies with higher SFRs tend to exhibit lower metallicities at fixed stellar mass, commonly interpreted as the dilution of the interstellar medium by metal-poor gas inflows \citep{Mannucci2010,Curti2020}. 
Whether the FMR is redshift-invariant or evolves with cosmic time remains actively debated. 
While some studies suggest that the FMR holds approximately up to $z\sim3$, others find significant deviations at higher redshifts (e.g. \citealt{Sanders2021,Heintz2023,Curti2024}, Westcott et al. in prep.). 
Recent \textit{JWST} observations indicate that galaxies at $z>4$ typically lie $\sim0.2$--0.3\,dex below the local FMR, consistent with elevated gas fractions and enhanced inflow rates in the early Universe \citep[e.g.][]{Nakajima2023,Fujimoto2024}.

Beyond internal galaxy properties, the MZR is increasingly recognised as sensitive to large-scale environment. 
Environmental processes such as ram-pressure stripping, strangulation, mergers, harassment, and pre-processing can significantly modify the gas supply and chemical evolution of galaxies \citep[e.g.][]{McCarthy2008,Peng2015,Boselli2022,Xu2025a,Brennan2015,Delahaye2017,Fujita2004,Werner2022,Lopes2024}. 
In the local Universe, galaxies in groups and clusters are observed to be modestly more metal enriched than field galaxies at fixed stellar mass, with typical offsets of $\sim0.05$--0.1\,dex \citep[e.g.][]{Peng2015,Darvish2015,Chartab2020}. 
Environmental effects on galaxy properties have also been detected out to $z\lesssim2$ in clusters and protoclusters \citep[e.g.][]{Hatch2017,Tadaki2019,Namiki2019,Wang2022,Lemaux2022,Liu2023,PerezMartinez2023,Forrest2024,Hughes2025}. 

At higher redshifts, however, the environmental dependence of the MZR remains uncertain and controversial. 
Some studies report metal-deficient galaxies in overdense regions at $z\sim2$--3, such as in the CL J1449+0856 cluster and several protoclusters \citep[e.g.][]{Valentino2015,Wang2022,PerezMartinez2024,Calabro2022,Sattari2021}, while others find evidence for metallicity enhancement in dense environments \citep[e.g.][]{Shimakawa2015,Shimakawa2018,PerezMartinez2023,Adachi2025}. 
In several cases, no significant environmental dependence has been detected \citep[e.g.][]{Namiki2019}, highlighting the difficulty of disentangling intrinsic physical effects from observational uncertainties, given that the reported offsets are often comparable to the measurement errors ($\sim0.1$\,dex). 
Hydrodynamical simulations further reveal a complex picture: environmental effects on metallicity may arise from both internal processes, such as cold-mode accretion and feedback-driven outflows, and external mechanisms, such as gas stripping, confinement, and recycling in dense environments \citep[e.g.][]{Wang2023b,Fukushima2023,Esposito2025,MorokumaMatsui2025}. 
Despite these advances, the impact of environment on galaxy chemical evolution within the first billion years after the Big Bang remains poorly constrained.

Theoretical models predict that environmental effects should emerge early. 
Cosmological simulations show that overdense regions collapse first, hosting halos with higher mass accretion rates, earlier star formation, and enhanced merger activity \citep{Chiang2017,Overzier2016}. 
Recent hydrodynamical simulations incorporating detailed feedback and metal transport further suggest that galaxies in dense environments may experience more efficient metal retention and accelerated chemical enrichment already at $z>5$ \citep[e.g.][]{Kannan2023,Garcia2024,Garcia2025,Dekel2023}. 
These predictions motivate a systematic observational test of environmental chemical evolution during the epoch of reionization.

Beyond metallicity scaling relations, environment may also regulate the global star-formation activity of the early Universe. 
Measurements of the cosmic star-formation rate density (SFRD) reveal a rapid rise from $z\sim10$ to $z\sim6$, followed by a decline toward lower redshifts \citep[e.g.][]{MadauDickinson2014,Harikane2022}. 
However, most existing measurements represent averages over all environments. 
If early star formation is preferentially concentrated in overdense regions, as predicted by theory, protoclusters may contribute disproportionately to the cosmic SFRD and play a critical role in shaping the topology and timing of cosmic reionization \citep[e.g.][]{Chiang2017,Wilkins2023}. 
Quantifying this environmental contribution is therefore essential for understanding both galaxy formation and the reionization history of the Universe \citep[e.g.][]{Qiong2024}.

In this work, we use deep photometric and spectroscopic \textit{JWST} observations from the CEERS and JADES surveys to investigate the environmental dependence of galaxy chemical enrichment and star formation at $z\sim4$--10. We characterise environment using the projected fifth-nearest-neighbour surface density, enabling a homogeneous separation of galaxies into underdense (field) and overdense (cluster-like) regions across both surveys. We combine robust stellar masses, SFRs, gas-phase metallicities, and structural measurements to address three key questions: (i) whether the MZR exhibits a measurable environmental dependence at $z>4$; (ii) how environment modulates the FMR and the coupling between metallicity and star formation; and (iii) whether overdense regions host an enhanced share of the cosmic star-formation rate density during the epoch of reionization.

This paper is structured as follows. In \S\ref{sec: data} we describe the CEERS and JADES data sets, sample selection, and environmental classification. \S\ref{sec: 3} presents the environmental dependence of the MZR and FMR, including Bayesian fits and residual analyses. In \S\ref{sec:4} we explore the connection between metallicity, galaxy size, and star-formation activity, and examine the evolution of the FMR relative to the local relation. \S\ref{sec:5} quantifies the environmental contribution to the cosmic SFRD at $z\sim6$--9 and compares our results with theoretical predictions. Finally, \S\ref{sec:6} summarises our conclusions and discusses implications for early galaxy formation and reionization.

Throughout this paper, we assume a flat cosmological model with $\Omega_{\Lambda} = 0.7, \Omega_{m} = 0.3$ and $H_0 = 70 \mkms \, \rm Mpc^{-1}$. All magnitudes used in this paper are in the AB system \citep{Oke1983}.

\section{Data and Sample Selection}\label{sec: data}

\subsection{Observations }\label{sec:data}

This work is based on deep imaging and spectroscopy from the \textit{James Webb Space Telescope} (\textit{JWST}) in the CEERS and JADES survey fields. 
The CEERS program (ID: 1345; PI: S. Finkelstein) provides multi-band \textit{JWST}/NIRCam imaging over the Extended Groth Strip (EGS), complemented by extensive \textit{HST} imaging from CANDELS \citep{Grogin2011,Koekemoer2011,Bagley2023}. 
The JADES program (IDs: 1180, 1210; PIs: Eisenstein \& Lützgendorf) delivers ultra-deep \textit{JWST}/NIRCam observations in the GOODS-S field, together with ancillary \textit{HST} data from the Hubble Legacy Fields \citep{Rieke2023}.

We adopt photometric catalogues constructed from the reduced CEERS and JADES imaging following the procedures described in the EPOCHS series \citep{Adams2024, Austin2025, Conselice2025, Harvey2025}. 
Source detection and photometry are performed on weighted NIRCam stacks, with aperture corrections derived from \texttt{WebbPSF} models \citep{Perrin2012,Perrin2014}. 

Photometric redshifts and stellar population parameters (stellar mass and SFR) are obtained via SED fitting with \texttt{Bagpipes} \citep{Carnall2018}. We perform SED fitting with \texttt{Bagpipes} using BC03 stellar population models \citep{Bruzual2003} and a Kroupa IMF \citep{kroupa2001MNRAS.322..231K}. 
We adopt a lognormal star formation history (SFH), parameterized by the cosmic time of peak star formation and its FWHM, with broad priors of $0.01$--$10$~Gyr \citep[e.g.][]{2023Natur.619..716C,2023MNRAS.519.5859W}. 
The SFH is truncated when exceeding the age of the Universe at the best-fit redshift. Nebular emission lines and continuum are included using \texttt{CLOUDY} models \citep{2013RMxAA..49..137F}, and dust attenuation is modelled with the \citet{calzetti2000} law. 
We adopt logarithmic priors for metallicity and dust attenuation, with metallicity in the range $[10^{-3}, 3] \, \text{Z}_{\odot}$, dust attenuation $A_V = [0.0001, 10]$, and ionization parameter $\log U = [-3, -1]$, consistent with expectations for young, metal-poor high-redshift galaxies.

In addition, we derive star formation rates independently from the dust-corrected rest-frame UV luminosity. Specifically, we fit a power-law to the observed photometry over the rest-frame $1250$--$3000\,\text{\AA}$ range to estimate the UV continuum slope, and apply the \citet{Meurer1999} relation to correct for dust attenuation at $1600\,\text{\AA}$. The dust-corrected UV luminosity is then converted into an SFR using the calibration of \citet{Kennicutt2012}. This UV-based SFR primarily traces the emission from short-lived massive stars and therefore reflects star formation averaged over a characteristic timescale of $\sim10$--$100$ Myr. While comparable in timescale to the SFRs inferred from SED fitting, this estimate is more directly tied to observables. While spectroscopic measurements are available for a subset of galaxies, we do not adopt them here as our primary SFR estimates due to their limited and heterogeneous coverage, which would introduce selection biases in the statistical analysis.

Spectroscopic redshifts and emission-line measurements are taken from publicly available CEERS, JADES and FRESCO NIRSpec catalogues where available \citep[e.g.][]{Nakajima2023,Curti2023,Curti2024,Fujimoto2024}. We obtain an outlier fraction of $\eta = 7.0\%$, defined by $|\Delta z|/(1+z_{\rm spec})>0.15$, and a normalized median absolute deviation of ${\rm NMAD}=1.48\,{\rm med}\,| (z_{\rm spec}-z_{\rm phot})/(1+z_{\rm spec}) |=0.038$, indicating good agreement between photometric and spectroscopic redshifts. 

\subsection{Gas-phase metallicity}\label{sec:metal}

In this work we adopt published gas-phase metallicities for our two fields, using the values provided by the corresponding spectroscopic analyses of the JADES and CEERS samples \citep{Curti2024,Nakajima2023}. Throughout, we take the gas-phase metallicity to be the oxygen abundance, expressed as $12+\log(\mathrm{O/H})$, and we use the quoted uncertainties from the original works.

For the JADES field, we adopt the metallicities derived by \citet{Curti2024}. In their analysis, NIRSpec PRISM ($R\sim100$) and medium-resolution gratings ($R\sim1000$) are treated independently, with emission-line fluxes measured (via \texttt{ppxf}, \citealt{Cappellari2017,Cappellari2023}) and corrected for dust attenuation using Balmer decrements when available. Specifically, \citet{Curti2024} use H$\alpha$/H$\beta$ at $z<6.75$ (where H$\alpha$ is covered by NIRSpec) and H$\gamma$/H$\beta$ at higher redshift, adopting a \citet{Gordon2003} attenuation law and Case~B recombination ratios at $T=1.5\times10^4$~K. When Balmer decrements cannot be measured, they employ the nebular attenuation inferred from SED fitting of PRISM spectra with \texttt{BEAGLE} \citep{ChevallardCharlot2016}, which self-consistently accounts for differential attenuation between stellar and nebular emission. Gas-phase metallicities are then inferred from a revised set of strong-line calibrations anchored to the $T_e$ abundance scale (building on \citealt{Curti2020}) and refined for the low-metallicity regime, and by combining multiple available diagnostics in an MCMC framework implemented with \texttt{emcee} \citep{ForemanMackey2013}. Given the limited detectability/coverage of [N\,{\sc ii}] and [S\,{\sc ii}] at these redshifts (and blending of [N\,{\sc ii}] with H$\alpha$ in PRISM spectra), the effective constraints for most sources primarily rely on ``$\alpha$-element''-based ratios (e.g.\ $R3$, $O32$, and, when available, [Ne\,{\sc iii}]/[O\,{\sc ii}]). \citet{Curti2024} also apply conservative quality controls, visually removing spectra with poor line fits and excluding AGN candidates (including both broad-line AGN and narrow-line candidates identified via high-ionisation features), to ensure applicability of star-forming strong-line calibrations \citep[e.g.][]{Maiolino2023,Scholtz2023}. When both PRISM and grating measurements are available, they prioritizes higher-resolution spectra, while leveraging the deeper PRISM spectra for fainter targets \citep{Curti2024}.

For the CEERS field, we adopt the $T_e$-based metallicities presented by \citet{Nakajima2023} for sources with significant detections of the auroral line [O\,{\sc iii}]$\lambda 4363$. In this direct method, the electron temperature of the O$^{2+}$ zone, $T_e(\mathrm{[O\,III]})$, is derived from the reddening-corrected [O\,{\sc iii}]$\lambda\lambda4363/5007$ ratio assuming a representative electron density (typically $n_e=100~\mathrm{cm^{-3}}$) and using \texttt{PyNeb} \citep{Luridiana2015}. The temperature in the O$^{+}$ zone, $T_e(\mathrm{[O\,II]})$, is then estimated from $T_e(\mathrm{[O\,III]})$ following standard prescriptions \citep{Izotov2006}. Ionic abundances are computed from the relevant line-to-H$\beta$ ratios, and the total oxygen abundance is obtained as $\mathrm{O/H}=\mathrm{O}^+/\mathrm{H}^+ + \mathrm{O}^{2+}/\mathrm{H}^+$, neglecting $\mathrm{O}^{3+}$ under the assumption of no strong He\,{\sc ii}$\lambda4686$ emission \citep{Nakajima2023,Izotov2006}. For cases where [O\,{\sc ii}] is not detected, \citet{Nakajima2023} assess the impact of the upper limits and show it is sub-dominant compared to the measurement uncertainties. While \citet{Nakajima2023} further discuss how strong-line indicators can be biased by ionization state at $z\gtrsim4$ and emphasize the importance of ionization corrections when using empirical calibrations \citep[e.g.][]{Nakajima2022}, in this work, we rely on their reported direct-method metallicities for the CEERS subsample.

%JADES: Curti2024
%CEERS: Nakajima2023

\subsection{Cluster galaxies selection}
\label{sec:environment}

To investigate the environmental dependence of galaxy chemical properties at high redshift, we construct samples of galaxies residing in low- and high-density regions using the CEERS and JADES datasets. These two JWST survey fields provide deep multi-wavelength imaging and spectroscopy over independent sightlines, allowing us to probe galaxy populations across a wide range of cosmic environments and redshifts.

\begin{figure} % plot_dist_z.py
    \centering
    \includegraphics[width=\columnwidth]{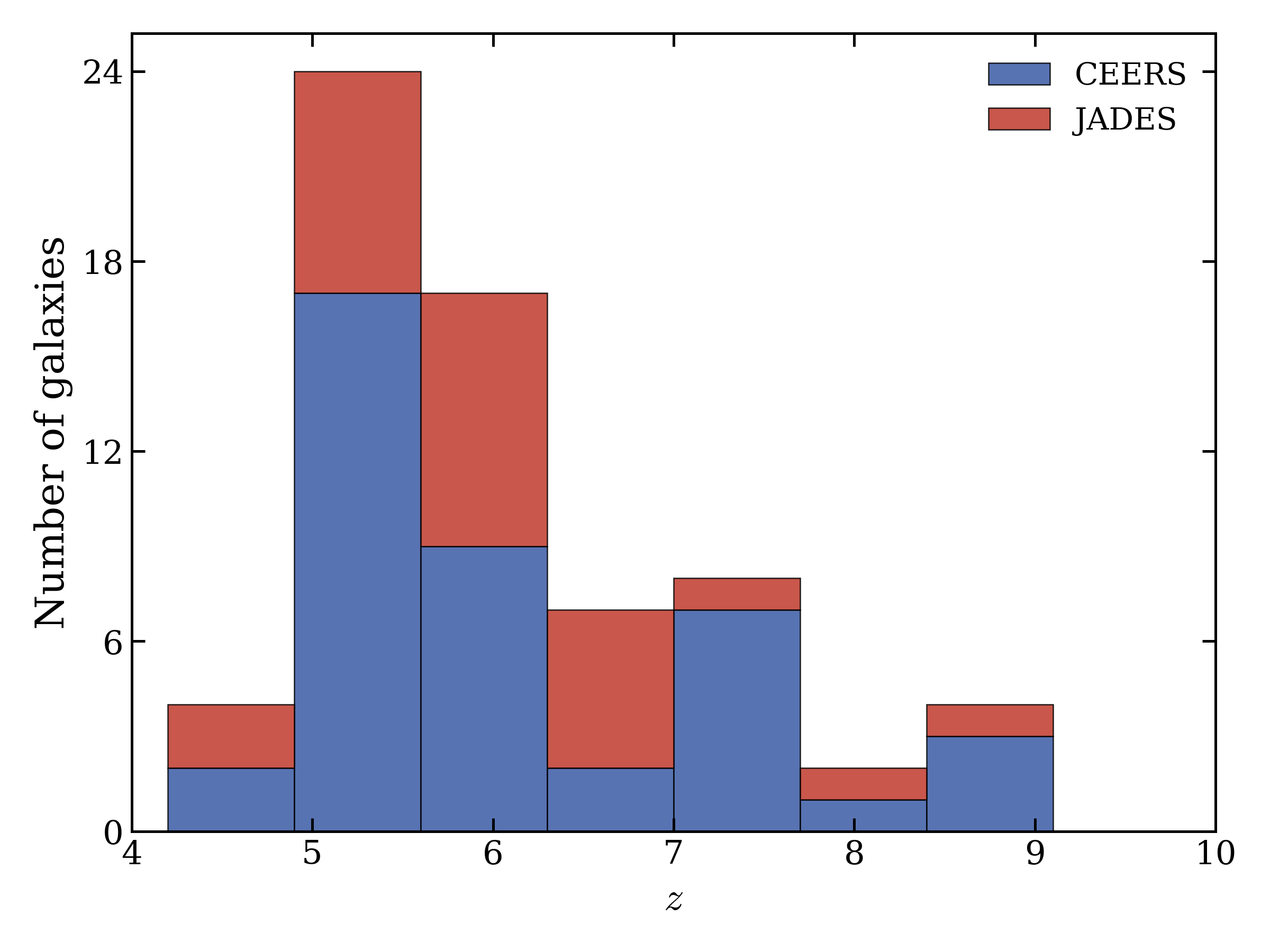}
    \caption{Spectroscopic redshift distributions of galaxies in the CEERS and JADES samples used in this work. The stacked histograms show the number of galaxies as a function of redshift for the two JWST fields, with CEERS shown in blue and JADES in red.}
    \label{fig:redshift_dist}
\end{figure}

Figure~\ref{fig:redshift_dist} shows the redshift distributions of galaxies in the CEERS and JADES spectroscopically confirmed samples used in this work at $z \simeq 4- 10$.

Firstly, we use the photometric data to quantify the local environment of each galaxy using the projected overdensity parameter $\Sigma_5$, which measures the surface density of galaxies within a fixed neighbourhood defined by the fifth nearest neighbour. We compute the projected fifth-nearest neighbor surface density $\Sigma_5$ in physical Mpc$^{-2}$ using galaxies within $\Delta z=\pm0.25$ of each target, for more details see \citet{Qiong2024}. This estimator has been widely adopted in studies of galaxy environments at both low and high redshift, as it provides a robust characterization of local density while minimizing sensitivity to shot noise in sparse samples. For each survey field, we define a threshold overdensity $\Sigma_{5,\mathrm{th}}$ to separate galaxies into low- and high-density subsamples. Specifically, in each redshift bin we compute the $\Sigma_5$ distribution separately for CEERS and JADES, and classify galaxies as {underdense} (field-like) if $\Sigma_5$ is below the 25th percentile, and {overdense} (protocluster-like) if $\Sigma_5$ is above the 75th percentile. 
Galaxies in the intermediate percentiles are not used in the primary environment comparison. In practice, the threshold is chosen according to the characteristic overdensity distribution of each field, ensuring that the resulting subsamples are statistically comparable. We emphasise that this definition distinguishes relative environmental density within each redshift slice rather than identifying extreme overdensities only. We also verified that our main conclusions are qualitatively unchanged when varying the percentile cuts (e.g. 20\%/80\%) and when adopting a single cut (e.g. 50\%).

The final cluster and field samples are constructed by requiring reliable measurements of stellar mass, gas-phase metallicity, and their associated uncertainties. This selection ensures that the derived environmental trends are not affected by poorly constrained physical parameters. After applying these criteria, we obtain well-defined samples of 27 galaxies in the low-density (field) environment and 29 galaxies in the high-density (protocluster) environment, which form the basis of the subsequent analysis of the mass–metallicity relation and the age–metallicity relation.

We note that the definition of cluster galaxies at $z \gtrsim 5$ inevitably involves uncertainties associated with projection effects, redshift errors, and the evolving nature of large-scale structure. However, the consistency of the environmental classification across two independent JWST fields, combined with the broad redshift overlap shown in Figure~\ref{fig:redshift_dist}, indicates that our results are robust against such systematic effects. In the following sections, we use these samples to explore how galaxy chemical enrichment depends on environment during the early stages of cosmic structure formation.

\section{Mass metallicity relation}\label{sec: 3}
In this section, we present the stellar mass--metallicity relation and the fundamental metallicity relation at $z=4$--10 using deep JWST observations in the CEERS and JADES fields. 
These relations provide a compact empirical summary of how galaxies regulate gas accretion, star formation, and metal retention. JWST now enables such tests for statistically meaningful samples at $z>4$. 
In this work, they are used to determine whether galaxies in denser regions are already chemically more mature, and whether any environmental imprint is amplified once star formation is taken into account. 
Throughout this section, environment is quantified using the projected fifth-nearest-neighbour surface density $\Sigma_5$, and galaxies are divided into \emph{low-density} and \emph{high-density} subsamples using a redshift-slice, field-specific percentile split (Section~\ref{sec:environment}).

\subsection{Environmental dependence of the mass--metallicity relation}\label{sec:3.1}

To assess whether the gas-phase metallicity of galaxies depends on their large-scale environment, we performed a Bayesian comparison of the mass--metallicity relation between galaxies in overdense (``cluster'') and low-density (``field'') regions. We use the combined CEERS and JADES samples (with consistent photometric/spectroscopic quality cuts; Section~\ref{sec:data}), yielding $N_{\rm low}=$27 and $N_{\rm high}=$29 galaxies in the two density bins.

\begin{figure} % mz.py
    \centering
    \includegraphics[width=\columnwidth]{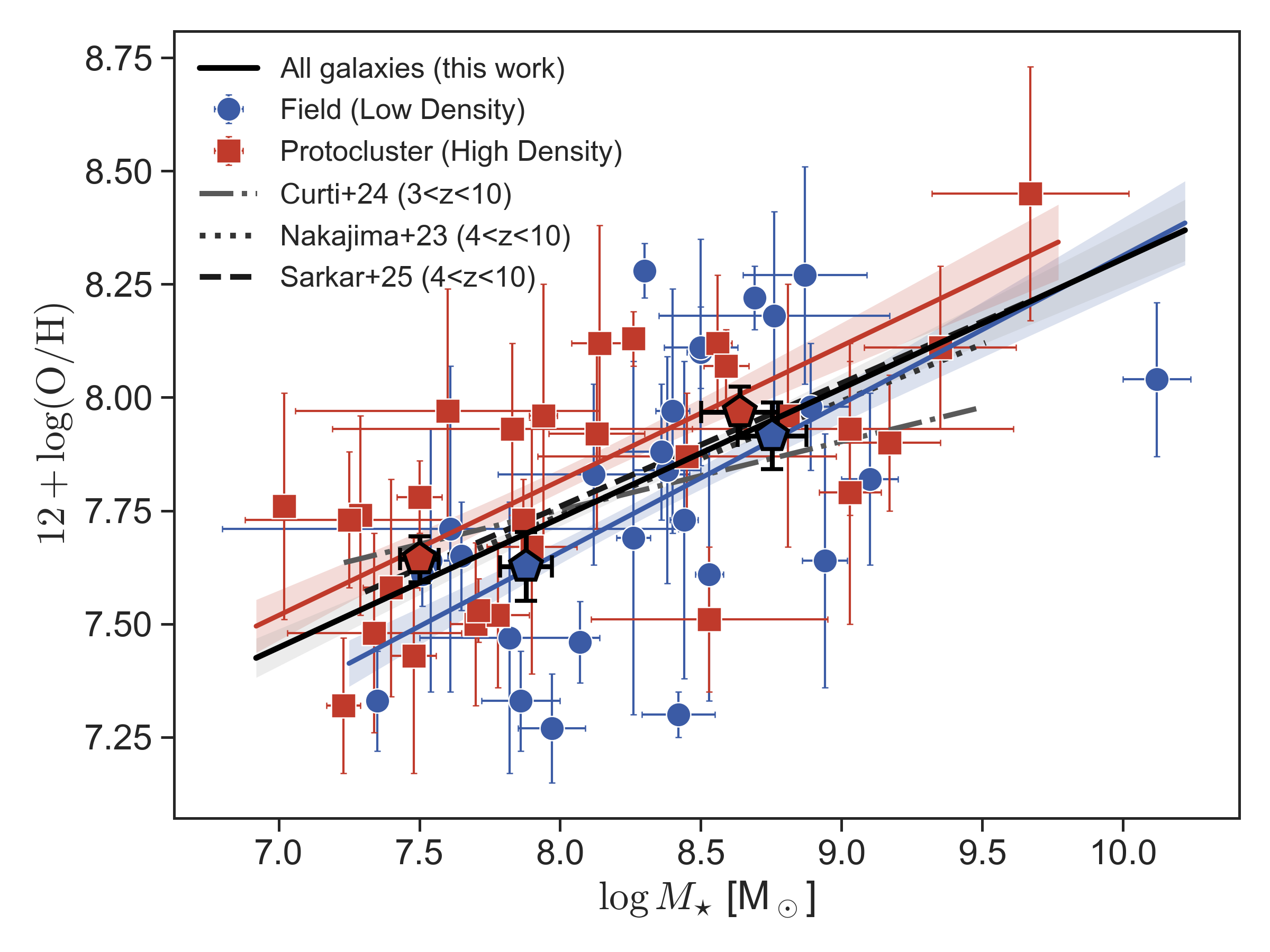}
    \caption{Mass--metallicity relation for field (blue) and cluster (red) galaxies at $z\sim5$--$7$. 
    Solid lines show the median posterior fits from our MCMC analysis, and points represent individual galaxies with measurement uncertainties. Dashed lines indicate representative relations from recent \textit{JWST} studies \citep[e.g.][]{Nakajima2023,Curti2024,Sarkar2025}.}
    \label{fig:mzr_fit}
\end{figure}

We modelled the relation between stellar mass and oxygen abundance as a linear form,
\begin{equation}
12 + \log(\mathrm{O/H}) = a (\log M_\star - M_0) + b,
\end{equation}
where $M_0$ is the median stellar mass of the combined sample, introduced to reduce the covariance between slope and intercept. 
For each environment, the posterior distributions of $(a,b,\ln\sigma_{\rm int})$ were inferred using an \texttt{emcee}-based MCMC sampler \citep{ForemanMackey2013}, assuming Gaussian errors that include both measurement uncertainties and an intrinsic scatter $\sigma_{\rm int}$.
To ensure robustness against outliers and sampling variance, we implemented a hierarchical bootstrap scheme: for each of 200 bootstrap resamplings of the data, we performed an ``inner'' MCMC fit after $3\sigma$ clipping of residuals. 
This procedure propagates both measurement and sampling uncertainties into the final posterior of the intercept difference $\Delta b = b_{\mathrm{cluster}} - b_{\mathrm{field}}$.

Figure~\ref{fig:mzr_fit} shows the resulting MZR fits for the two environments. 
Cluster galaxies (red) tend to occupy the upper envelope of the relation compared to field galaxies (blue). A linear fit yields:
\begin{align*}
\text{Field:} & \quad y = (0.33 \pm 0.05)x + (5.04 \pm 0.37), \\
\text{Cluster:} & \quad y = (0.30 \pm 0.05)x + (5.44 \pm 0.40),
\end{align*}
where $x = \log M_\star / M_\odot$ and $y = 12 + \log({\rm O/H})$.
Cluster galaxies exhibit a $\sim0.1$ dex metallicity excess at fixed mass, consistent with earlier hints of accelerated chemical evolution in dense regions. This modest difference suggests a possible environmental modulation of early metal enrichment.

\begin{figure} % mz_mcmc.py
    \centering
    \includegraphics[width=\columnwidth]{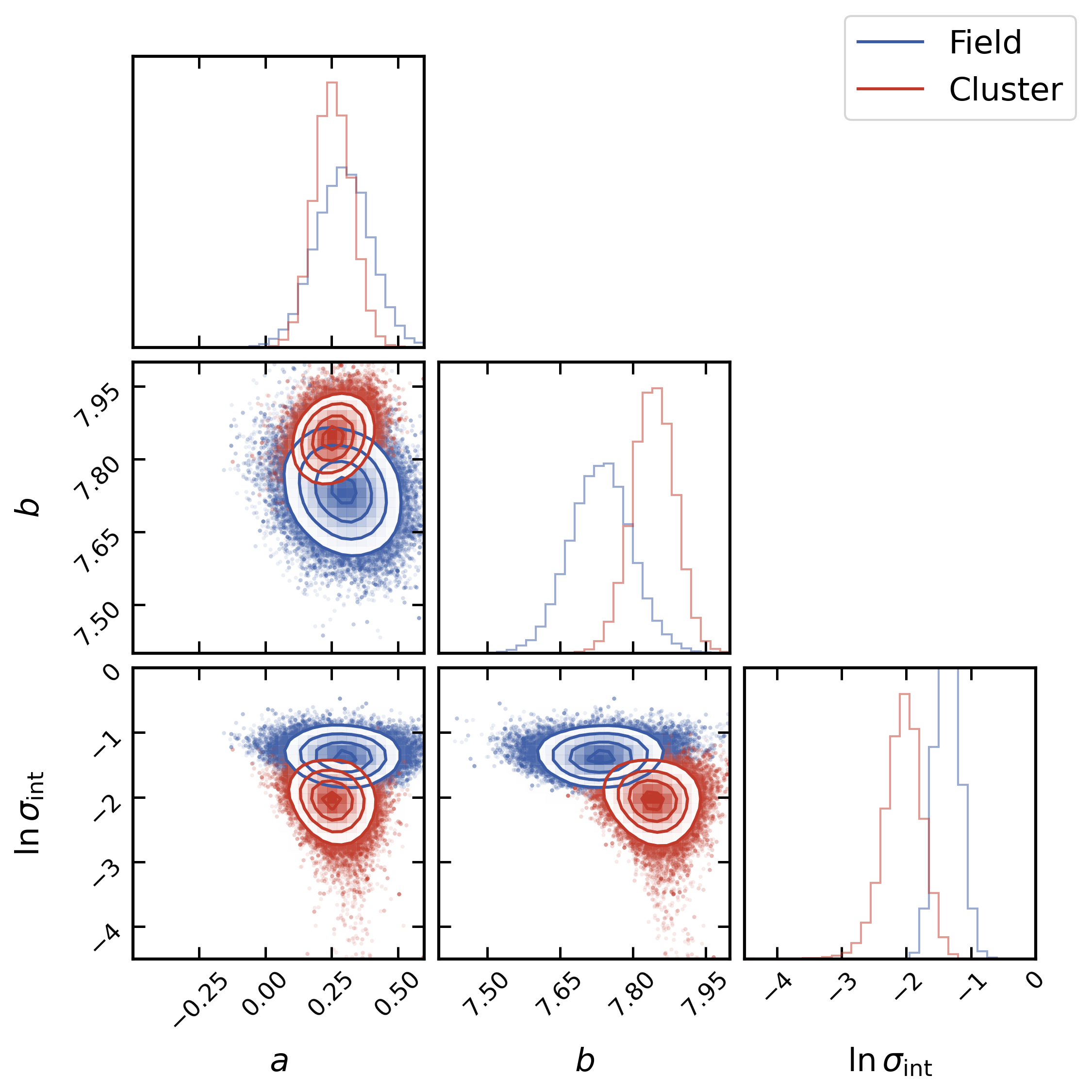}
    \caption{Joint posterior distributions of slope $a$, intercept $b$, and intrinsic scatter $\ln\sigma_{\rm int}$ for field (blue) and cluster (red) galaxies. 
    Cluster galaxies exhibit a slightly higher intercept at fixed slope, consistent with a mild metallicity enhancement.}
    \label{fig:corner_mzr}
\end{figure}

\begin{figure}
    \centering
    \includegraphics[width=0.48\textwidth]{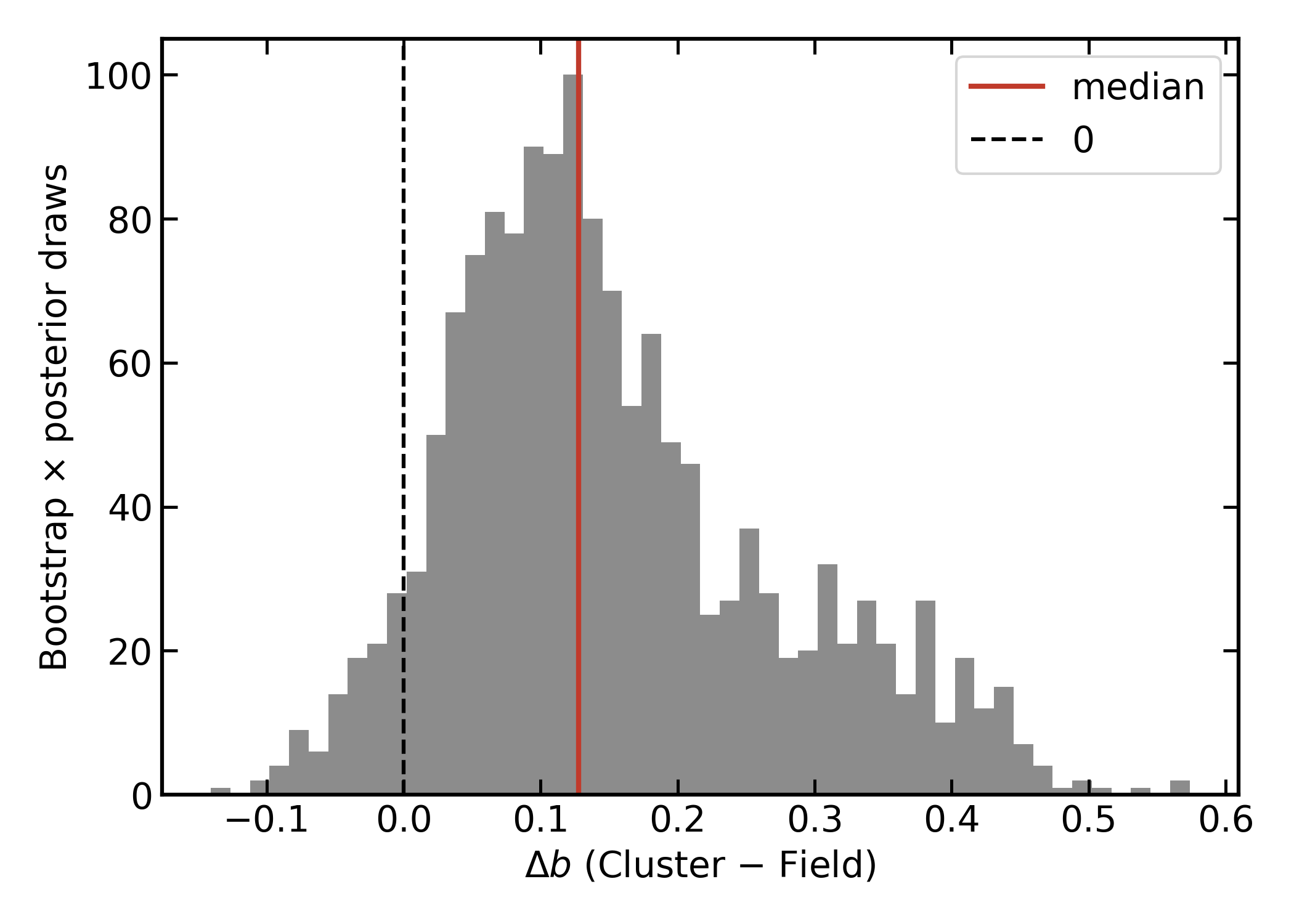}
    \caption{Bootstrap--MCMC posterior distribution of the intercept difference $\Delta b = b_{\mathrm{cluster}} - b_{\mathrm{field}}$ after $3\sigma$ clipping. 
    The median ($\Delta b=0.14$ dex) and 95\% credible interval are indicated by the red line and shaded region, respectively.}
    \label{fig:delta_b}
\end{figure}

The joint posterior distributions of the MZR parameters for both samples are displayed in Figure~\ref{fig:corner_mzr}, revealing well-constrained slopes ($a\simeq0.26$) and a systematic offset in intercept.
The bootstrap--MCMC posterior of the intercept difference is shown in Figure~\ref{fig:delta_b}. 
We find a median offset of $
\Delta b = 0.14^{+0.26}_{-0.21}~{\rm dex}$,
with a 95\% credible interval of $[-0.07,\,0.40]$ and a posterior probability $\Pr(\Delta b > 0) = 0.89$. 
This indicates that cluster galaxies are moderately ($\approx0.1$--$0.2$~dex) more metal enriched than field galaxies at fixed stellar mass, with the significance exceeding the conventional $0.68$ threshold.

Our results are consistent with growing evidence that galaxies in overdense regions at $z\gtrsim5$ tend to exhibit more advanced chemical enrichment than their field counterparts. Recent \textit{JWST} studies reveal substantial diversity in metallicity at high redshift and suggest that early-forming systems may reach higher metal abundances at fixed stellar mass \citep[e.g.][]{Nakajima2023,Curti2024,Fujimoto2024,Lizihao2025}. At lower redshifts ($z\lesssim3$), modest metallicity enhancements of $\sim0.05$--$0.1$~dex in group and cluster environments have been reported \citep[e.g.][]{Peng2015,Darvish2015,Chartab2020,Hatch2017,Tadaki2019}, indicating that the MZR is systematically modulated by environment across cosmic time. Our measurements extend this trend into the epoch of reionization. While the present offset remains modest given the uncertainties, the trend aligns with theoretical predictions that galaxies in dense environments experience earlier halo assembly, more efficient metal retention, and enhanced gas recycling \citep[e.g.][]{Chiang2017,Kannan2023,Dekel2023,Garcia2024,Garcia2025}.

\subsection{Environmental dependence of the mass--metallicity--SFR relation}

While the mass--metallicity relation provides a first-order description of chemical enrichment, it does not fully capture the role of star formation in regulating galaxy metallicity. We therefore extend our analysis by incorporating star formation rate and exploring the mass--metallicity--SFR relation in different environments, with the aim of assessing whether environmental effects persist in the three-dimensional parameter space of $M_\star$, SFR, and metallicity. This approach allows us to test whether overdense and underdense galaxies follow distinct fundamental metallicity relations at $z\sim4$--10.

\subsubsection{The fundamental metallicity relation}

\begin{figure}
\centering
\includegraphics[width=\columnwidth]{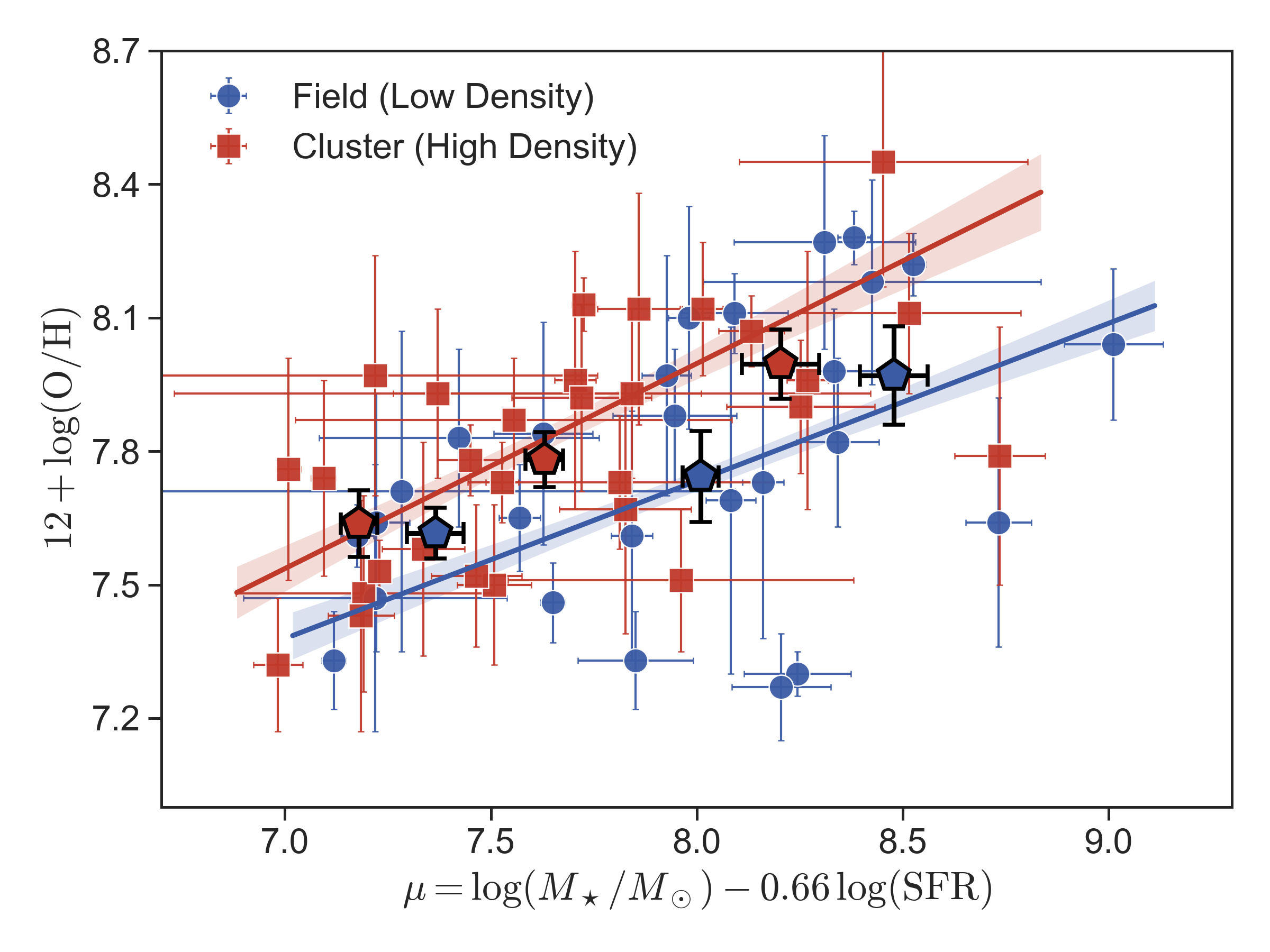}
\caption{FMR for CEERS and JADES galaxies. Solid lines show the median posterior fits,
and points represent individual galaxies with measurement uncertainties. Cluster galaxies show on average higher metallicities at fixed $\mu$.}
\label{fig:fmr}
\end{figure}

Figure~\ref{fig:fmr} presents the gas-phase metallicity as a function of the FMR parameter 
$\mu = \log M_\star - \alpha \log \mathrm{SFR}$, with $\alpha = 0.66$ adopted from \citet{Mannucci2010}. Here, the SFR is derived from dust-corrected rest-frame UV luminosity and therefore traces star formation averaged over a timescale of $\sim$10--100 Myr. Compared to the MZR, the scatter in metallicity is moderately reduced, indicating that the inclusion of star formation activity captures part of the intrinsic variance in chemical enrichment, consistent with previous studies of the FMR at both low and high redshift \citep{Curti2020,Sanders2021,Curti2024}. A linear fit in the $\mu$--metallicity plane yields:
\begin{align*}
\text{Field:} & \quad 12+\log({\rm O/H}) = (0.35 \pm 0.05)\,\mu + (4.90 \pm 0.33), \\
\text{Cluster:} & \quad 12+\log({\rm O/H}) = (0.46 \pm 0.07)\,\mu + (4.31 \pm 0.45),
\end{align*}
where uncertainties represent the 16th--84th percentile ranges from the posterior distributions. 
The two slopes differ by $\Delta a = 0.11$ ($\sim1.3\sigma$). Adopting $\mu\simeq8.0$, close to the median of our sample, the best-fit relations imply $\Delta Z(\mu=8.0) \equiv Z_{\rm cluster} - Z_{\rm field} \simeq 0.29~{\rm dex}$. Given the uncertainties in the fitted parameters, this offset is significant at the $\sim1$--$2\sigma$ level. The systematic metallicity enhancement of galaxies in overdense regions across the observed $\mu$ range suggests that environment modulates the coupling between stellar mass, star formation activity, and chemical enrichment.

This environmental offset is comparable to, though slightly larger than, the metallicity differences reported in group and cluster environments at $z\lesssim2$ \citep{Peng2015,Darvish2015,Chartab2020,Hatch2017}, and is broadly consistent with recent \textit{JWST} studies suggesting accelerated chemical enrichment in early overdensities \citep{Lizihao2025,Fujimoto2024}. 
However, the magnitude of the offset remains smaller than the intrinsic scatter of the relation. The marginally steeper slope observed in overdense regions suggests a stronger coupling between metallicity and star formation activity, potentially reflecting enhanced gas recycling efficiency or earlier gas consumption in dense environments \citep{Bassini2024,PerezDiaz2024,BakerMaiolino2023,DuartePuertas2022}.

\subsubsection{The distribution in the $M_\star$--SFR--$Z$ space}

To further characterise the environmental dependence of chemical enrichment beyond the $\mu$-based FMR, 
we examine the distribution of stellar mass, star formation rate, and gas-phase metallicity in the $M_\star$--SFR plane.
This representation provides a complementary view of how metallicity varies at fixed stellar mass and star formation activity in different environments.

\begin{figure} % FMR.py
    \centering
    \includegraphics[width=0.98\linewidth]{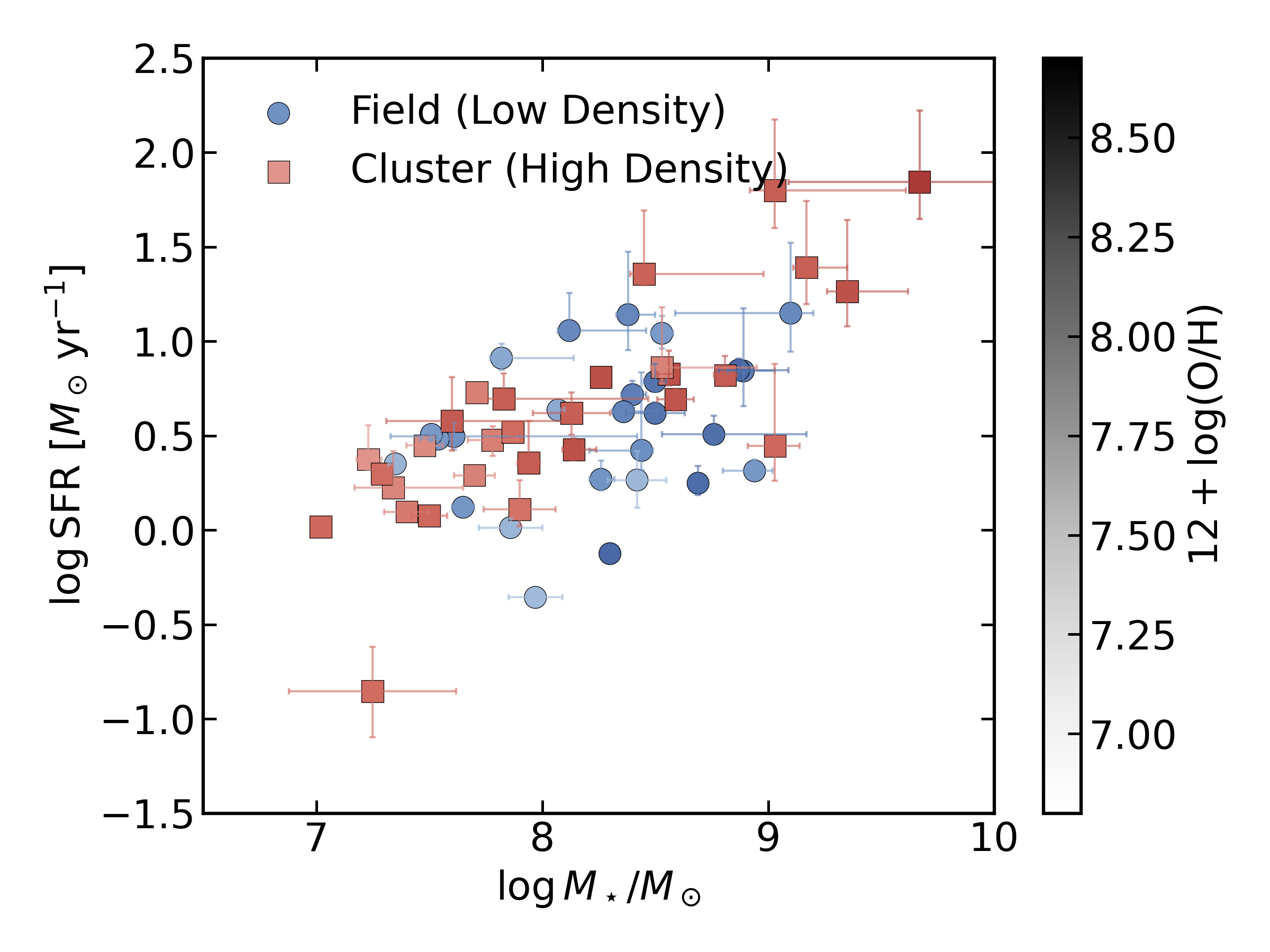}
    \caption{
        Fundamental metallicity relation in different environments at $z\sim4$--$10$.
        The distribution of galaxies from the CEERS and JADES samples in the 
        $\log(\mathrm{SFR})$--$\log(M_\star/M_\odot)$ plane, 
        color-coded by environment: blue points correspond to 
        low-density (field, $\Sigma_5 < P_{25,\Sigma_5}$) galaxies, while red points indicate 
        high-density (cluster,  $\Sigma_5 > P_{75,\Sigma_5}$) galaxies. 
        Individual error bars show the measurement uncertainties in stellar mass 
        and star formation rate. 
        The grayscale colorbar represents the gas-phase metallicity 
        ($12 + \log(\mathrm{O/H})$), with lighter shades indicating higher metallicity. 
        The figure illustrates that galaxies in dense environments tend to exhibit 
        higher metallicities at fixed stellar mass and SFR, 
        suggesting accelerated chemical enrichment in protocluster regions.
    }
    \label{fig:FMR_env_graybar}
\end{figure}

Figure~\ref{fig:FMR_env_graybar} presents the distribution of galaxies in the $\log(\mathrm{SFR})$--$\log(M_\star/M_\odot)$ plane, color-coded by environment. Field (low-density) galaxies are shown in blue, while protocluster (high-density) members are shown in red. Individual error bars indicate measurement uncertainties in both stellar mass and SFR. The grayscale color bar encodes the gas-phase metallicity, $12 + \log(\mathrm{O/H})$. At fixed $M_\star$ and SFR, galaxies in overdense regions populate systematically higher metallicities than field galaxies, with a typical offset of order $\sim0.2$--0.3 dex, suggesting that metal enrichment proceeds more rapidly in overdense regions \citep[e.g.,][]{Chartab2021,Bischetti2024}.

\begin{figure*} %fmz_smooth.py
    \centering
    \includegraphics[width=\textwidth]{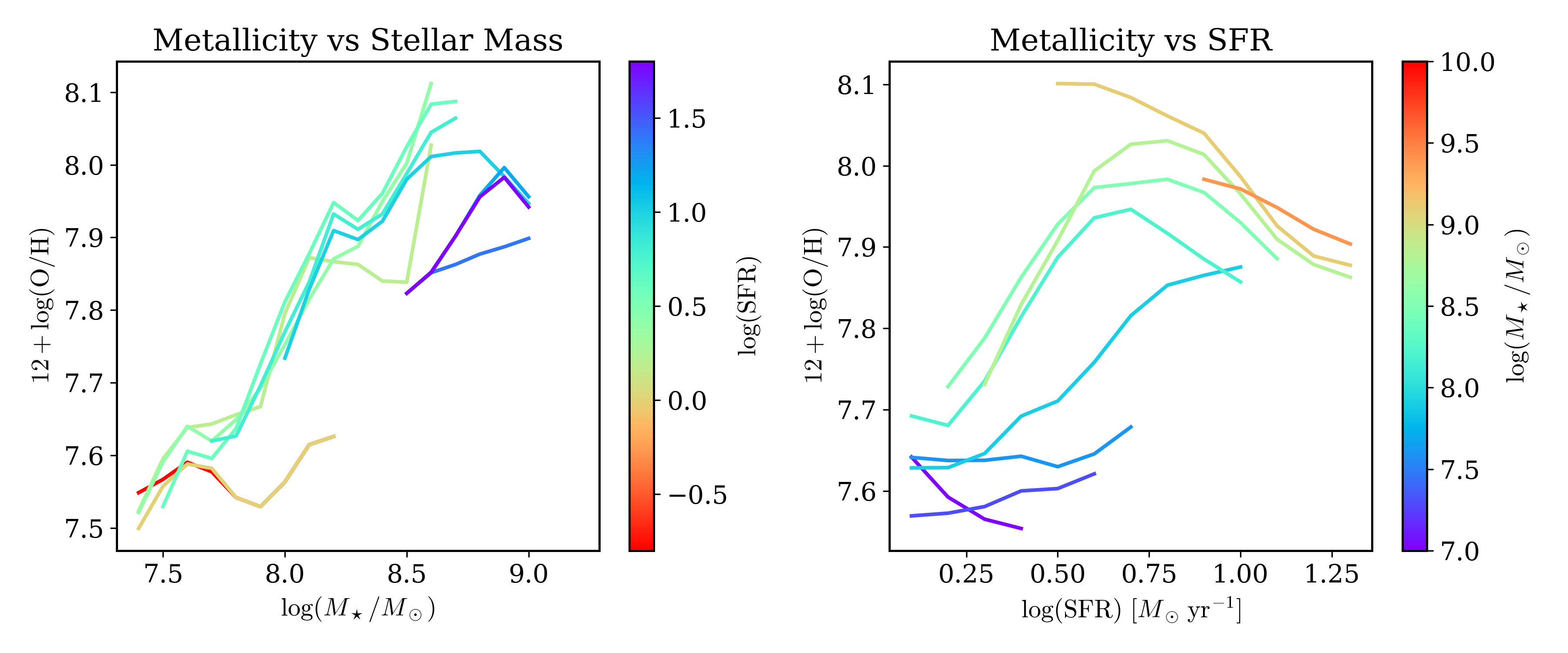}
    \caption{
        Left: Gas-phase metallicity ($12+\log(\mathrm{O/H})$) as a function of stellar mass, 
        color-coded by star formation rate (SFR) measured within a sliding window of width 
        $\Delta\log(\mathrm{SFR}) = 0.8$\,dex. 
        Right: The same metallicity as a function of $\log(\mathrm{SFR})$, 
        color-coded by stellar mass within a sliding window of width 
        $\Delta\log(M_\star/M_\odot) = 1.2$\,dex. 
        Curves represent the median metallicities computed in overlapping moving bins 
        ($\pm0.2$\,dex in $\log M_\star$ or $\pm0.3$\,dex in $\log\mathrm{SFR}$), 
        with mild Savitzky–Golay smoothing applied to reduce statistical fluctuations. 
        The color gradients indicate that, at fixed stellar mass, galaxies with lower SFRs 
        tend to be more metal-poor, consistent with the evolving form of the 
        fundamental metallicity relation seen in both CEERS and JADES samples.
    }
    \label{fig:FMR_slidingwindow}
\end{figure*}

To quantify the dependence of metallicity on mass and SFR, we apply a sliding-window median analysis. 
The left panel of Figure~\ref{fig:FMR_slidingwindow} shows the median $12 + \log(\mathrm{O/H})$ as a function of stellar mass in bins of SFR, while the right panel shows the dependence on SFR in bins of stellar mass. 
Both trends reveal that metallicity increases with $M_\star$ at fixed SFR, and decreases with increasing SFR at fixed mass, consistent with the existence of the qualitative form of ``fundamental metallicity relation'' observed up to $z \sim 3$ 
\citep[e.g.,][]{Sanders2018,Curti2020,Looser2024,Pistis2024}. 
However, the normalization of the relation at $z > 4$ is $\sim0.2$--0.3~dex lower than in the local Universe, reflecting the lower chemical maturity of early galaxies \citep[e.g.,][]{Nakajima2023,Curti2023,Fujimoto2024}. These results suggest that the coupling between stellar mass, star formation activity, and metallicity is already established by $z\sim8$, 
although with a normalization offset relative to the local Universe.

\subsubsection{The Fitting of the Fundamental Metallicity Plane}

To quantitatively assess the dependence of gas-phase metallicity on both stellar mass and star formation activity, 
we model the galaxies in our sample using the {fundamental metallicity relation}, expressed as a linear plane in logarithmic space:
\begin{equation}
12 + \log(\mathrm{O/H}) = a \, \log(M_\star/M_\odot) + b \, \log(\mathrm{SFR}) + c,
\end{equation}
where $a$ describes the sensitivity of metallicity to stellar mass growth (i.e.\ self-enrichment in a closed-box system), 
$b$ represents the inverse sensitivity to star formation rate (reflecting the efficiency of gas inflow and dilution), 
and $c$ is the normalization constant associated with the overall metal retention efficiency.

We fit this plane separately for galaxies in the low-density (field) and high-density (cluster) environments using a 
Markov Chain Monte Carlo (MCMC) approach implemented with the \texttt{emcee} ensemble sampler 
\citep{ForemanMackey2013}. The likelihood function accounts for both measurement uncertainties and an intrinsic 
scatter term $\sigma$, such that
\begin{equation}
\ln \mathcal{L} = -\frac{1}{2} \sum_i 
\left[ \frac{(Z_i - a M_i - b S_i - c)^2}{\sigma^2 + \delta Z_i^2} 
+ \ln(2\pi(\sigma^2 + \delta Z_i^2)) \right],
\end{equation}
where $Z_i = 12+\log(\mathrm{O/H})$, $M_i = \log(M_{\star,i}/M_\odot)$, and $S_i = \log(\mathrm{SFR}_i)$. We use flat priors within physically motivated ranges ($-1.5<a<2.5$, $-2.5<b<1.5$, $6<c<10$, $0<\sigma<1.5$), 
and sample 4000 steps with 64 walkers after a burn-in phase of 1500 steps. The best-fit parameters are derived from 
the median of the posterior distributions, with uncertainties corresponding to the 16th and 84th percentiles. The best-fit coefficients are summarised in Table~\ref{tab:fmr}. 

\begin{table}
\centering
\caption{Best-fit parameters of the FMR plane for field and cluster galaxies. The quoted uncertainties represent statistical errors only; systematic uncertainties from metallicity calibration and SED fitting are likely at the level of 0.1–0.2 dex.}
\label{tab:fmr}
\begin{tabular}{lcccc}
\hline
 & $a$ & $b$ & $c$ & $\sigma$ [dex] \\
\hline
Field   & $0.278\pm0.001$ & $+0.016\pm0.001$ & $5.449\pm0.001$ & $0.090$ \\
Cluster & $0.261\pm0.001$ & $-0.031\pm0.001$ & $5.722\pm0.001$ & $0.054$ \\
\hline
\end{tabular}
\end{table}

The stellar-mass dependence, quantified by the coefficient $a$, is consistent between the two environments within the uncertainties, 
with $a \simeq 0.27$ in both samples. This indicates that the primary scaling between stellar mass and metallicity remains broadly similar in low- and high-density regions. In contrast, the coefficient associated with star formation rate, $b$, exhibits a systematic difference between environments. Cluster galaxies show a negative correlation between metallicity and SFR ($b \simeq -0.03$), whereas field galaxies display a much weaker dependence ($b \simeq 0.02$). Although the statistical significance of this difference is modest, it suggests that the coupling between star formation activity and chemical enrichment may vary with environment. We further find that the intrinsic scatter of the relation is smaller in overdense regions ($\sigma \simeq 0.05$ dex) than in the field ($\sigma \simeq 0.09$ dex). This trend, while subject to substantial uncertainties, is consistent with a scenario in which galaxies in dense environments follow a more tightly regulated evolutionary pathway.

\subsubsection{Weak correlation between metallicity residuals and SFR or environment}

To further test whether star formation activity or environment introduces secondary trends beyond the primary mass dependence, 
we examine the metallicity residuals defined as 
\begin{equation}
\Delta Z = (12+\log(\mathrm{O/H})) - f(M_\star),
\end{equation}
where $f(M_\star)$ represents the baseline mass--metallicity relation. The function $f(M_\star)$ is derived from the combined CEERS+JADES sample using a non-parametric locally weighted regression (LOWESS; \citealt{Cleveland1979}) of $12+\log(\mathrm{O/H})$ as a function of $\log(M_\star/M_\odot)$. 
We adopt a smoothing fraction of ${\rm frac}=0.4$, which captures the global MZR while suppressing small-scale fluctuations. 
To assess the robustness of the baseline relation, we repeat the LOWESS fitting with 200 bootstrap resamplings and estimate the associated $68\%$ confidence interval.

% test_sfr_s5.py
We applied Spearman’s rank correlation coefficient ($\rho$)  to quantify the residuals $\Delta Z$ 
against $\log(\mathrm{SFR})$ and $\log(\Sigma_5)$, where $\Sigma_5$ is the projected distance 
to the fifth nearest neighbour. 
No significant correlation was found in either the field or cluster environments 
($|\rho| < 0.15$, $p > 0.4$).
In the field, $\Delta Z$ shows a weak positive correlation with $\log(\mathrm{SFR})$ 
($\rho \simeq +0.08$), while the cluster sample shows a similarly weak negative trend 
($\rho \simeq -0.03$).
For the full sample, we find 
$\rho_{\Delta Z-\log\mathrm{SFR}} = +0.02$ and 
$\rho_{\Delta Z-\log \Sigma_5} = +0.12$, 
corresponding to less than $1\sigma$ significance.
These results indicate that, after removing the dominant dependence on stellar mass, any residual correlations of metallicity with star formation rate or environment are weak and remain below the sensitivity of the current data. 
This does not contradict the environmental trends inferred from the FMR plane, but rather suggests that environmental effects act as a second-order modulation on top of the primary mass-driven relation.

The absence of a strong $\Delta Z$--SFR anti-correlation contrasts with the classical FMR observed in the local Universe \citep{Mannucci2010,Curti2020,Sanders2021}, 
but is broadly consistent with recent \textit{JWST} studies indicating large intrinsic scatter and stochastic gas accretion in high-redshift galaxies \citep[e.g.,][]{Nakajima2023,Curti2024,Fujimoto2024}. 
Similarly, the weak dependence on $\Sigma_5$ suggests that environmental regulation of chemical enrichment remains subtle at $z>4$, 
in agreement with theoretical models in which early galaxies are primarily governed by rapid gas inflows and bursty star formation, with environmental effects emerging gradually over cosmic time \citep[e.g.,][]{Chiang2017,Kannan2023,Dekel2023,Garcia2024}.

We verified that our conclusions are insensitive to the choice of smoothing scale in the LOWESS fitting. 
Varying the smoothing fraction from ${\rm frac}=0.3$ to $0.5$, or excluding the lowest and highest $5\%$ of the stellar-mass range to mitigate edge effects, does not alter the results within the statistical uncertainties.

\section{Environmental Acceleration of Galaxy Growth}\label{sec:4}

\subsection{Redshift Evolution of the Fundamental Metallicity Relation at \ensuremath{z>4}}

To quantify the redshift evolution of the fundamental metallicity relation (FMR), we compare our CEERS \citep{Nakajima2023} and JADES \citep{Curti2024} galaxies to the local \citet{Andrews2013} calibration. \citet{Andrews2013} used direct $T_e$ metallicities for local low-mass galaxies and showed that adopting $\alpha=0.66$ minimises the scatter in the $\mu_\alpha$--metallicity plane. A fair comparison of $\Delta Z$ across different high-redshift studies requires controlling for the adopted $\alpha$ in $\mu_\alpha$, and systematic differences among strong-line metallicity calibrations. Recent work has noted that some commonly used strong-line calibrations may yield metallicities that are systematically higher than $T_e$-anchored scales by $\sim$0.1--0.2 dex at low metallicity, potentially biasing inferred FMR offsets at $z>4$ (e.g. \citealt{Chakraborty2025,Lizihao2025}). In what follows, we therefore adopt the \citet{Andrews2013} $\mu_{0.66}$ formalism consistently for our sample and for literature comparisons, and we interpret any absolute $\Delta Z$ values with the caveat that residual calibration systematics at the $\sim$0.1--0.2 dex level may remain. 
We do not apply an IMF conversion between Chabrier and Kroupa, since the difference is small compared to the metallicity uncertainties and has a negligible impact on the inferred offsets for our purposes \citep{madau2014cosmic,Nakajima2023}. For each galaxy, we compute the deviation from the local FMR as
\begin{equation}
\Delta Z_{\rm FMR} = [12 + \log(\mathrm{O/H})]_{\rm obs} - [12 + \log(\mathrm{O/H})]_{\rm FMR}.
\end{equation}

\begin{figure} % MRZ_difference.py
    \centering
    \includegraphics[width=\columnwidth]{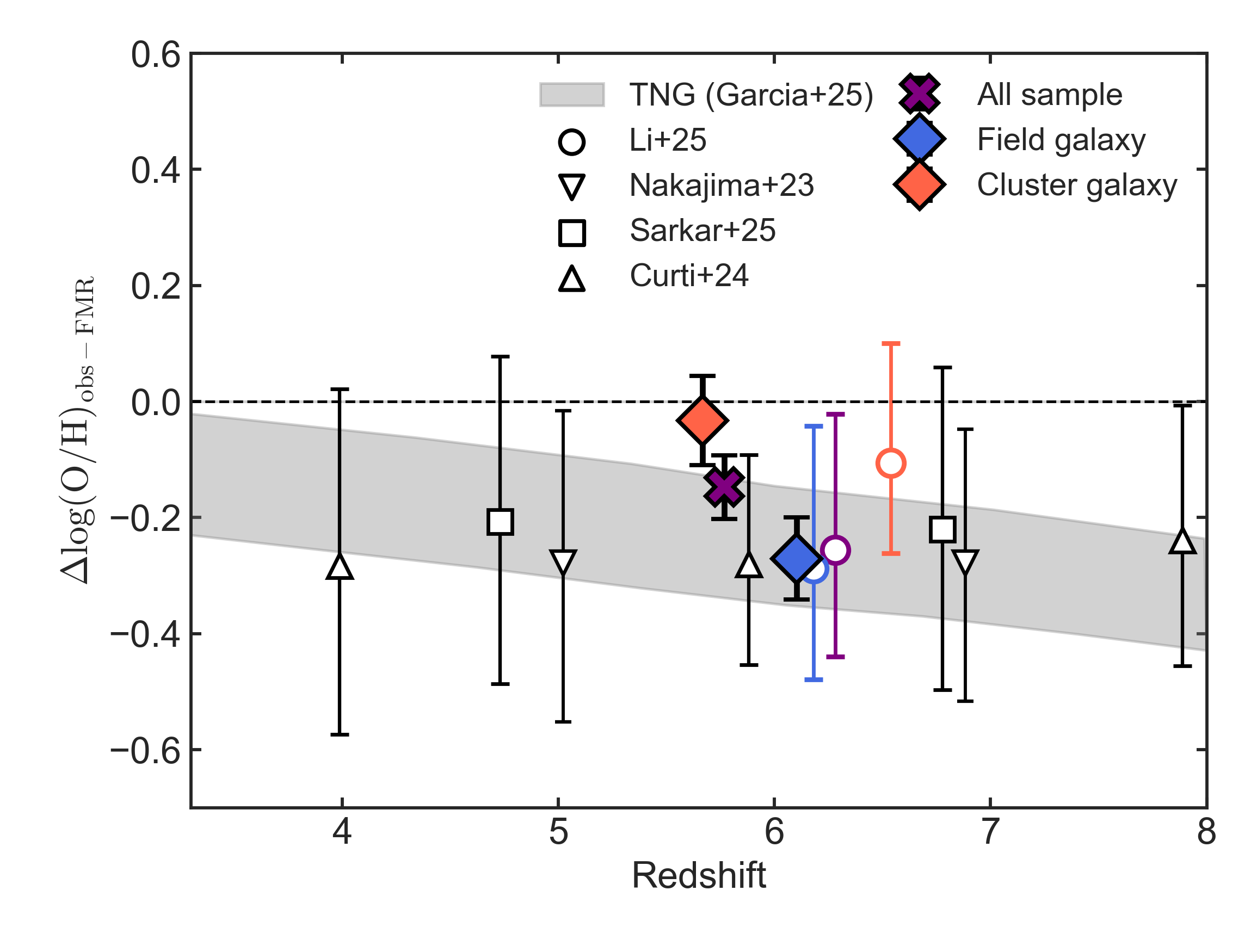}
    \caption{
    Offset from the local FMR of \citet{Andrews2013} as a function of redshift. 
    Filled points show our CEERS and JADES galaxies in different environments, while open symbols denote literature data. 
    The gray shaded region represents the \textsc{TNG} simulation predictions from \citet{Garcia2025}. 
    Our galaxies show a moderate metallicity deficit relative to the local FMR, consistent with previous results and indicative of reduced chemical enrichment efficiency at high redshift.
    }
    \label{fig:FMR_evolution}
\end{figure}

\noindent Figure~\ref{fig:FMR_evolution} shows the resulting offsets $\Delta Z$ as a function of redshift, compared with literature data and predictions from hydrodynamical \textsc{TNG} simulations \citep{Garcia2025}. We find that our full sample (purple) lies on average $\Delta Z \simeq -0.25$ dex below the local \citet{Andrews2013} relation, in broad agreement with the metal dilution reported at $z\gtrsim4$--6 in recent JWST studies (e.g. \citealt{Heintz2024,Curti2024}). Splitting by environment, galaxies in overdense regions show a smaller deficit (i.e. higher metallicity at fixed $\mu_{0.66}$) than the underdense subsample, consistent with accelerated enrichment in early overdensities. This is in line with the mild environmental dependence predicted in recent hydrodynamical simulations (e.g. \citealt{Garcia2024,Garcia2025}), which also suggest that the effective coupling between metallicity and SFR may evolve with redshift.

At similar redshifts, literature measurements exhibit a substantial scatter in the inferred FMR offsets. \citet{Curti2024} report relatively modest deviations from the local FMR ($\Delta Z \sim -0.2$\,dex) up to $z \simeq 8$, whereas \citet{Nakajima2023} and \citet{Sarkar2025} find significantly larger offsets ($\Delta Z \gtrsim -0.5$\,dex). Such discrepancies are likely driven by a combination of systematic effects, including differences in metallicity calibrations, emission-line diagnostics, and the adopted FMR parameterizations. In particular, as discussed by \citet{Lizihao2025}, both the choice of the slope parameter $\alpha$ in the $\mu_\alpha$ formalism and the selection of strong-line calibrations can introduce systematic shifts in metallicity estimates at the level of $\sim0.1$--0.2\,dex. 
At $z>4$, additional uncertainties may arise from evolving ionization conditions, abundance ratios, and selection effects, further amplifying the apparent dispersion in $\Delta Z$ across different studies. For this reason, we restrict our comparison to studies whose metallicity calibrations and FMR formalism are broadly consistent with our adopted framework. 
Measurements showing extreme offsets are not included in Fig.~\ref{fig:FMR_evolution}, as they are likely dominated by methodological systematics rather than intrinsic physical differences.

We find that our galaxies lie systematically below the local FMR, indicating lower chemical enrichment efficiency at early cosmic times. 
This offset is likely driven by the higher gas fractions and enhanced inflow of metal-poor material in the early Universe, as also suggested by \citet{Garcia2024,Garcia2025}. 
The environmental dependence suggests that these baryon-cycling processes are already operating by $z\sim6$, while overdense regions reach higher chemical maturity earlier than the field.

\subsection{Metallicity–Size Relation}

To investigate how chemical enrichment connects with the structural growth of galaxies, 
we derived the effective radius ($R_e$) for each source by fitting a single Sérsic profile 
to its two-dimensional light distribution. 
We used \texttt{GALFIT} version~3.0.5 \citep{Galfit1, Galfit2}, 
a two-dimensional least-squares fitting code that minimizes the reduced chi-squared statistic 
($\chi_{\mathrm{red}}^2$) to determine the best-fitting structural parameters. The point-spread function (PSF) models were generated using \texttt{WebbPSF} \citep{Perrin2014} and resampled to match the pixel scale of the science images, following the procedure described in \citet{Westcott2025}.
The Sérsic model describes the surface brightness profile as

\begin{equation}
    I(R) = I_e \exp \left\{ -b_n 
    \left[ \left( \frac{R}{R_e} \right)^{1/n} - 1 \right] \right\},
    \label{eq:sersic}
\end{equation}
where $I(R)$ is the intensity at radius $R$ from the galaxy center,
$R_e$ is the effective (half-light) radius enclosing half of the total flux, 
and $I_e$ is the intensity at $R_e$. 
The Sérsic index $n$ quantifies the concentration of the light profile:
$n \simeq 1$ corresponds to an exponential disk, while $n \simeq 4$ represents a de~Vaucouleurs bulge 
\citep{Sersic, Ciotti1991}. 
The coefficient $b_n$ depends on $n$ and can be approximated as 
$b_n \approx 2n - 0.327$ for $1 < n < 10$ \citep{ciotti1999}.

For each galaxy, we selected the \textit{JWST}/NIRCam filter that most closely corresponds 
to the rest-frame optical wavelength at its redshift, thereby minimizing the effects of 
morphological $k$-correction \citep{TaylorMager2007}.
Neighboring objects and bright residuals were masked iteratively to ensure stable fits, 
and sky background levels were either fixed or fitted depending on the local environment.
The uncertainties on $R_e$ and $n$ were estimated from the covariance matrix of the 
$\chi^2$ minimization and further validated by injecting artificial sources of comparable 
brightness into the science frames.

\begin{figure}
    \centering

    % ---------- (a) Z–Re ----------
    \begin{subfigure}[t]{\columnwidth}
        \centering
        \includegraphics[width=\columnwidth]{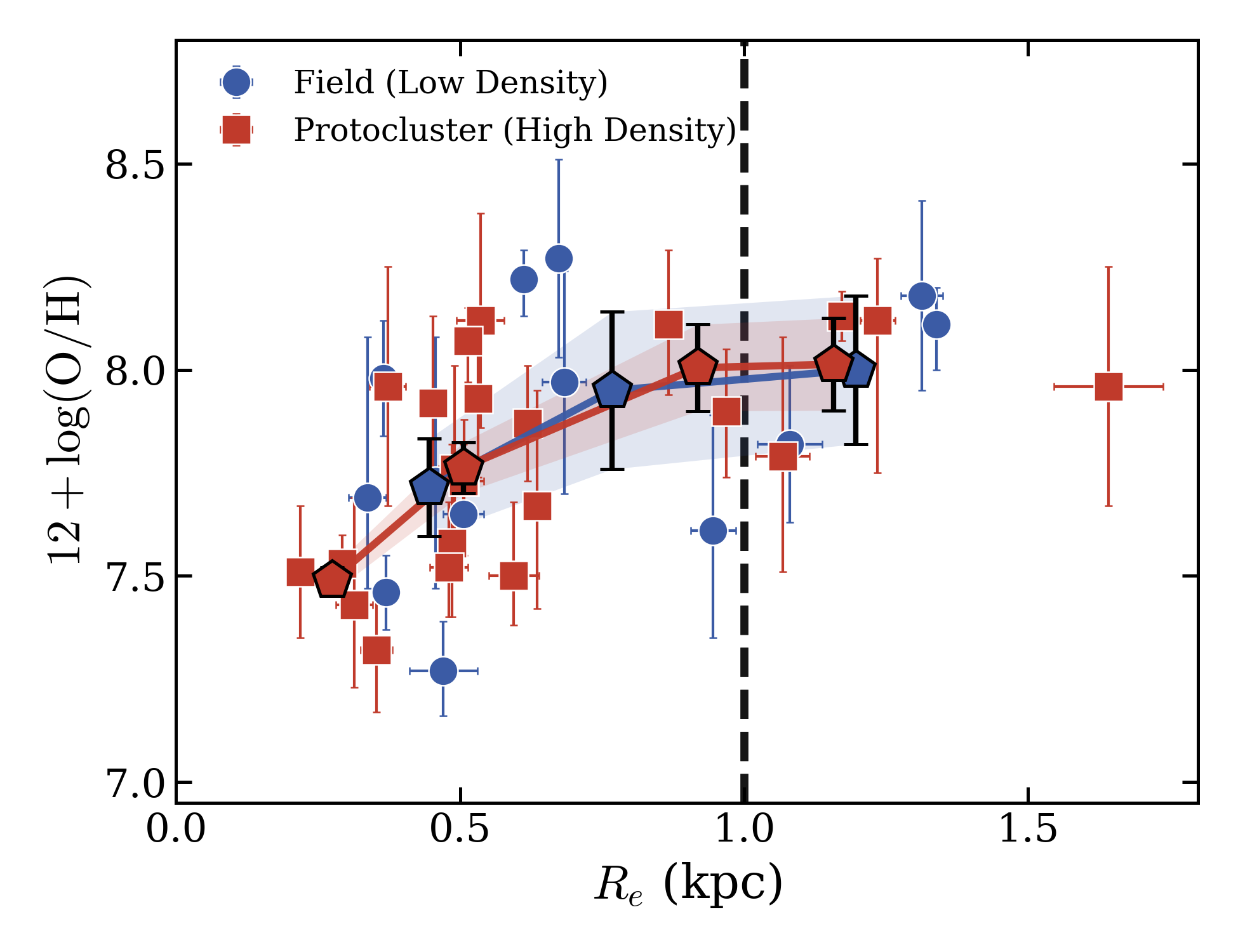} % Re.py
        \caption{
        Gas-phase metallicity as a function of effective radius ($R_e$) for galaxies in different environments.
        Individual galaxies are shown as faint points with measurement uncertainties, while solid lines indicate binned averages with $1\sigma$ errors on the mean.
        Field (low-density) and protocluster (high-density) galaxies are shown in blue and red, respectively.
        }
        \label{fig:Z_Re_profiles}
    \end{subfigure}

    \vspace{0.6em} % 调整两图间距

    % ---------- (b) ΔZ–Re ----------
    \begin{subfigure}[t]{\columnwidth}
        \centering
        \includegraphics[width=\columnwidth]{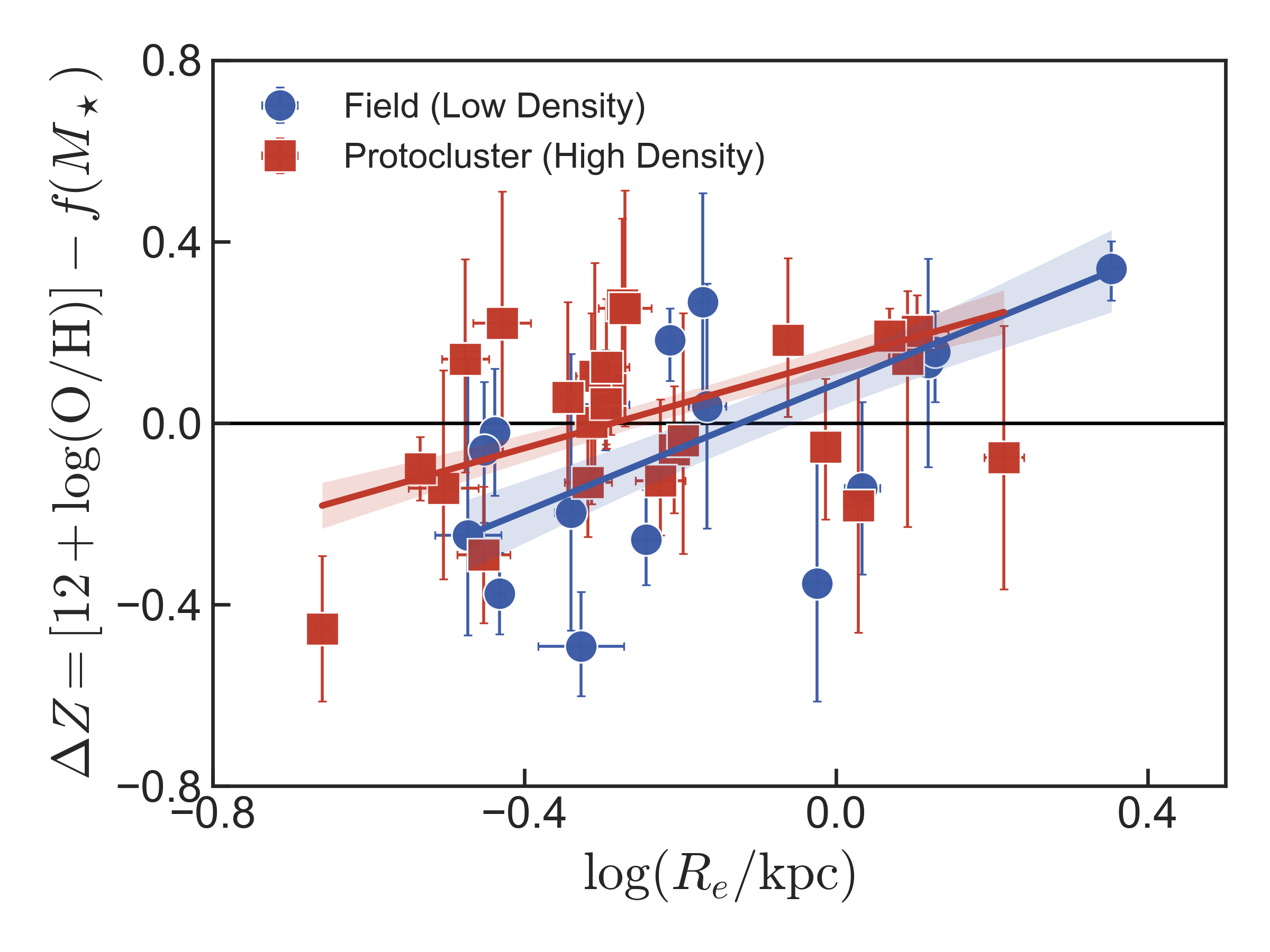} %test_re.py
        \caption{
        Residual metallicity as a function of galaxy effective radius for field (blue circles) and protocluster (red squares) galaxies.
        The residual metallicity, defined as 
        $\Delta Z = [12 + \log(\mathrm{O/H})] - f(M_\star)$,
        is measured relative to the global mass--metallicity relation derived from the combined CEERS and JADES samples.
        Error bars represent the uncertainties in both $R_e$ and gas-phase metallicity.
        Solid lines show the best-fitting linear trends for each environment.
        }
        \label{fig:MSR_residual}
    \end{subfigure}

    \caption{
    \textbf{(a)}~Metallicity--size relation and 
    \textbf{(b)}~mass-controlled residual metallicity--size relation for CEERS and JADES galaxies.
    }
    \label{fig:MSR_combined}
\end{figure}

We then combined the derived structural parameters with the gas-phase metallicities 
($12 + \log(\mathrm{O/H})$) measured from our CEERS and JADES samples 
to explore the relation between chemical abundance and galaxy size. Figure~\ref{fig:Z_Re_profiles} shows the relation between gas-phase metallicity and effective radius ($R_e$) for galaxies in both field and protocluster environments.
A weak but systematic trend is observed where metallicity increases with galaxy size up to $\sim1$~kpc. The correlation strength is modest, with Spearman coefficients of $\rho \sim 0.2$--0.3.
Such a correlation may reflect the combined effects of extended star-formation regions and efficient mixing in more spatially distributed systems \citep[e.g.,][]{Belfiore2017,Sharda2021,Curti2023}.

Beyond $\sim1$~kpc, the metallicity appears to flatten, implying that chemical enrichment saturates once the bulk of the gas reservoir has been processed.
While both field and protocluster galaxies follow similar overall trends, the field sample exhibits a slightly steeper metallicity–size relation, consistent with a stronger dependence of chemical abundance on internal structure.
This difference may indicate that, in dense environments, feedback and gas inflow are more tightly regulated, reducing the sensitivity of metallicity to galaxy morphology.
Overall, the results suggest that the metallicity–size relation at high redshift is mild and likely secondary to the dominant dependence on stellar mass.

%{\color{red}To investigate whether galaxy size plays an independent role in chemical enrichment, we examine the relation between gas-phase metallicity and effective radius after removing the primary dependence on stellar mass, which is known to strongly correlate with metallicity through the mass–metallicity relation.} For each galaxy, we compute the residual metallicity
%\begin{equation}
%\Delta Z = [12 + \log(\mathrm{O/H})] - f(M_\star),
%\end{equation}
%where $f(M_\star)$ represents the global mass–metallicity trend derived from a non-parametric LOWESS fit to the full sample. This subtraction effectively controls for the strong stellar-mass dependence of metallicity, allowing us to isolate any structural dependence.

To investigate whether galaxy size plays an independent role in chemical enrichment, we examine the relation between gas-phase metallicity and effective radius after removing the primary dependence on stellar mass, which is known to strongly correlate with metallicity through the mass--metallicity relation. 
For each galaxy, we subtract the global mass--metallicity trend, $f(M_\star)$, from the measured metallicity, $12 + \log(\mathrm{O/H})$. Here, $f(M_\star)$ is derived from a non-parametric LOWESS fit to the full sample. This subtraction effectively controls for the strong stellar-mass dependence of metallicity, allowing us to isolate any structural dependence.

In Figure~\ref{fig:MSR_residual}, we then explore the correlation between $\Delta Z$ and the logarithmic effective radius, $\log R_e$. A moderate positive correlation is found in the field sample ($\rho = 0.52$, $p = 0.04$, $\simeq2\sigma$), suggesting that, at fixed mass, larger galaxies tend to be slightly more metal-rich. 
Such a trend may reflect more extended star-formation histories, where larger systems continue forming stars over longer times and gradually build up metals. It may also be that gas in more extended galaxies is better mixed, so metals are more evenly distributed and less affected by local dilution \citep[e.g.,][]{Belfiore2017,Boardman2021}. In contrast, the protocluster sample exhibits a weaker, statistically insignificant correlation ($\rho = 0.29$, $p = 0.16$), implying that metallicity in dense environments is less sensitive to structural differences. These results suggest that the metallicity–size relation is largely driven by the mutual correlation of both metallicity and size with stellar mass, with galaxy size introducing at most a weak secondary modulation.

This environmental contrast indicates that chemical enrichment in the field remains more susceptible to processes related to gas accretion and spatially extended star formation, while galaxies in overdense regions appear to evolve toward more self-regulated, closed systems where size no longer dictates gas-phase metallicity \citep{Sharda2021}. The overall weak $\Delta Z$–$R_e$ dependence suggests that the apparent metallicity–size trend largely originates from the mutual correlations of both parameters with stellar mass, rather than from a direct structural effect.

\begin{figure} % Re_v1.py
    \centering
    \includegraphics[width=\columnwidth]{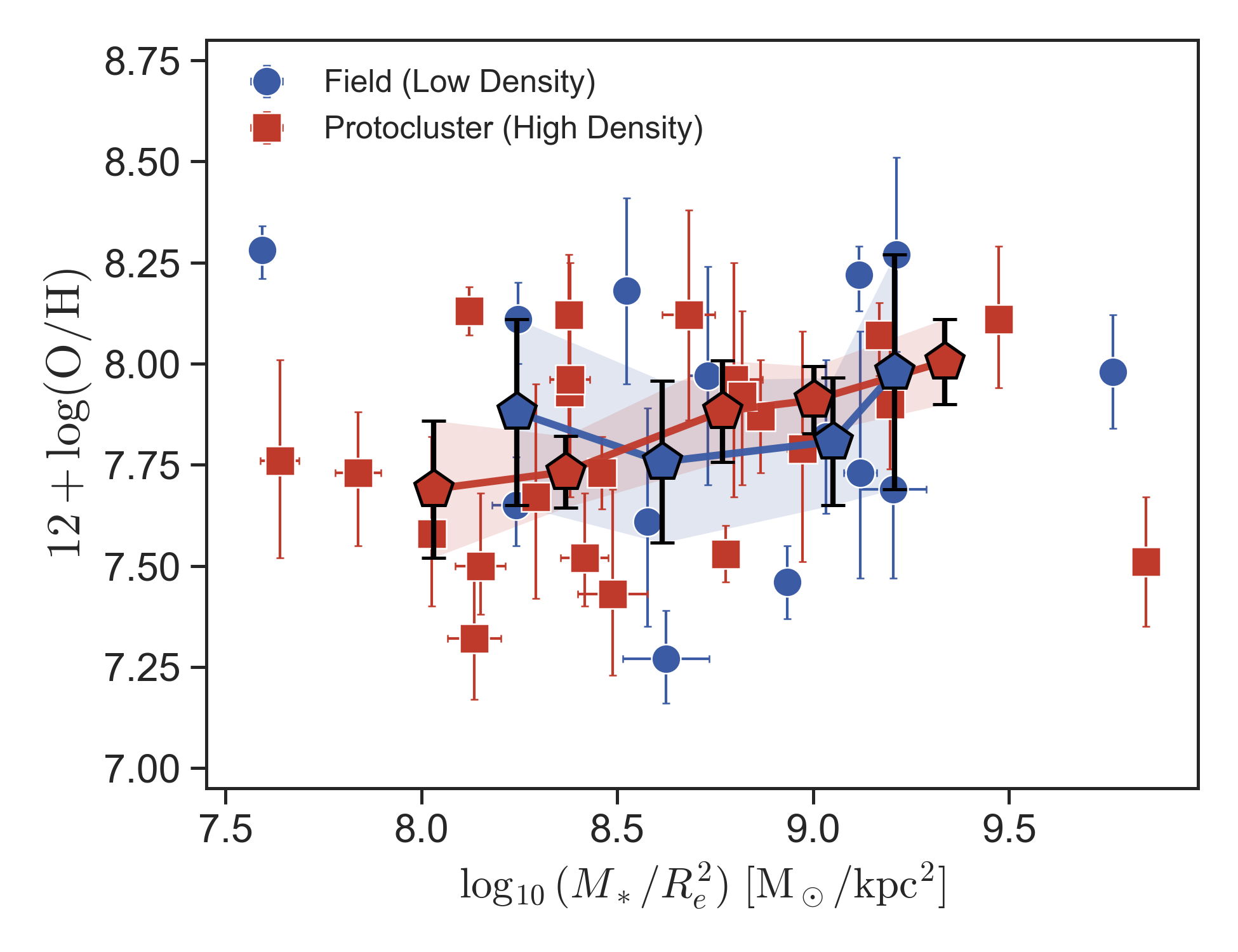}
    \caption{Gas-phase metallicity as a function of the stellar mass surface-density $\log_{10}(M_\ast/R_e^2)$ for galaxies.
    Blue circles show field (low-density) galaxies and red squares show protocluster
    (high-density) galaxies.
    Error bars include uncertainties in $12+\log(\mathrm{O/H})$ and the propagated
    uncertainties in $\log_{10}(M_\ast/R_e^2)$.
    Large pentagons mark binned means in $\log_{10}(M_\ast/R_e^2)$ with the shaded bands
    indicating the standard error on the mean.}
    \label{fig:Z_MRe2_z4to6p5}
\end{figure}

Figure~\ref{fig:Z_MRe2_z4to6p5} shows the gas-phase metallicity,
$12+\log(\mathrm{O/H})$, as a function of the stellar mass surface-density $\log_{10}(M_\ast/R_e^2)$ for galaxies in 
field (low density) and protocluster (high density). The larger pentagon symbols indicate binned means computed in uniform bins of $\log_{10}(M_\ast/R_e^2)$; the shaded bands represent the standard error on the mean within each bin. Overall, we find a positive correlation between metallicity and $\log_{10}(M_\ast/R_e^2)$, consistent with the expectation that more compact systems at fixed higher stellar surface density tend to be more chemically enriched. But the field and protocluster relations are broadly consistent within the measurement uncertainties and the bin-to-bin scatter, with no compelling evidence for a large environment-driven offset in $12+\log(\mathrm{O/H})$ at fixed
$\log_{10}(M_\ast/R_e^2)$.

\subsection{Metallicity Age Relation}

We explore the relation between stellar population age and gas-phase metallicity for galaxies at $5<z<10$ in the CEERS and JADES fields. Stellar population parameters, including mass-weighted ages and their uncertainties, are derived from SED fitting, while gas-phase metallicities are inferred from rest-frame optical emission-line diagnostics enabled by JWST/NIRSpec. The age--metallicity relation provides a complementary perspective to the MZR and FMR, linking the temporal buildup of stellar populations to chemical enrichment and offering insight into the timescales of baryon cycling in the early Universe \citep[e.g.,][]{Maiolino2008,Lilly2013,SomervilleDave2015}.

To quantify the age--metallicity relation, we model the dependence of metallicity on stellar age in logarithmic space as
\begin{equation}
12 + \log(\mathrm{O/H}) = a \log_{10}\left(\frac{t_{\mathrm{mw}}}{\mathrm{Gyr}}\right) + b,
\end{equation}
where $t_{\mathrm{mw}}$ denotes the mass-weighted stellar age derived from our \textsc{Bagpipes} fitting. We perform orthogonal distance regression (ODR) to account for measurement uncertainties in both age and metallicity. Uncertainties in $\log_{10}(t_{\mathrm{mw}})$ are obtained via error propagation from the linear age uncertainties. To assess the robustness of the fitted parameters, we further perform bootstrap resampling with 2000 realizations, in which galaxies are resampled with replacement and the ODR fit is repeated. The 16th--84th percentiles of the resulting parameter distributions are adopted as confidence intervals, and the bootstrap realizations are used to construct confidence bands for the fitted relations.

\begin{figure}
    \centering
    \includegraphics[width=\columnwidth]{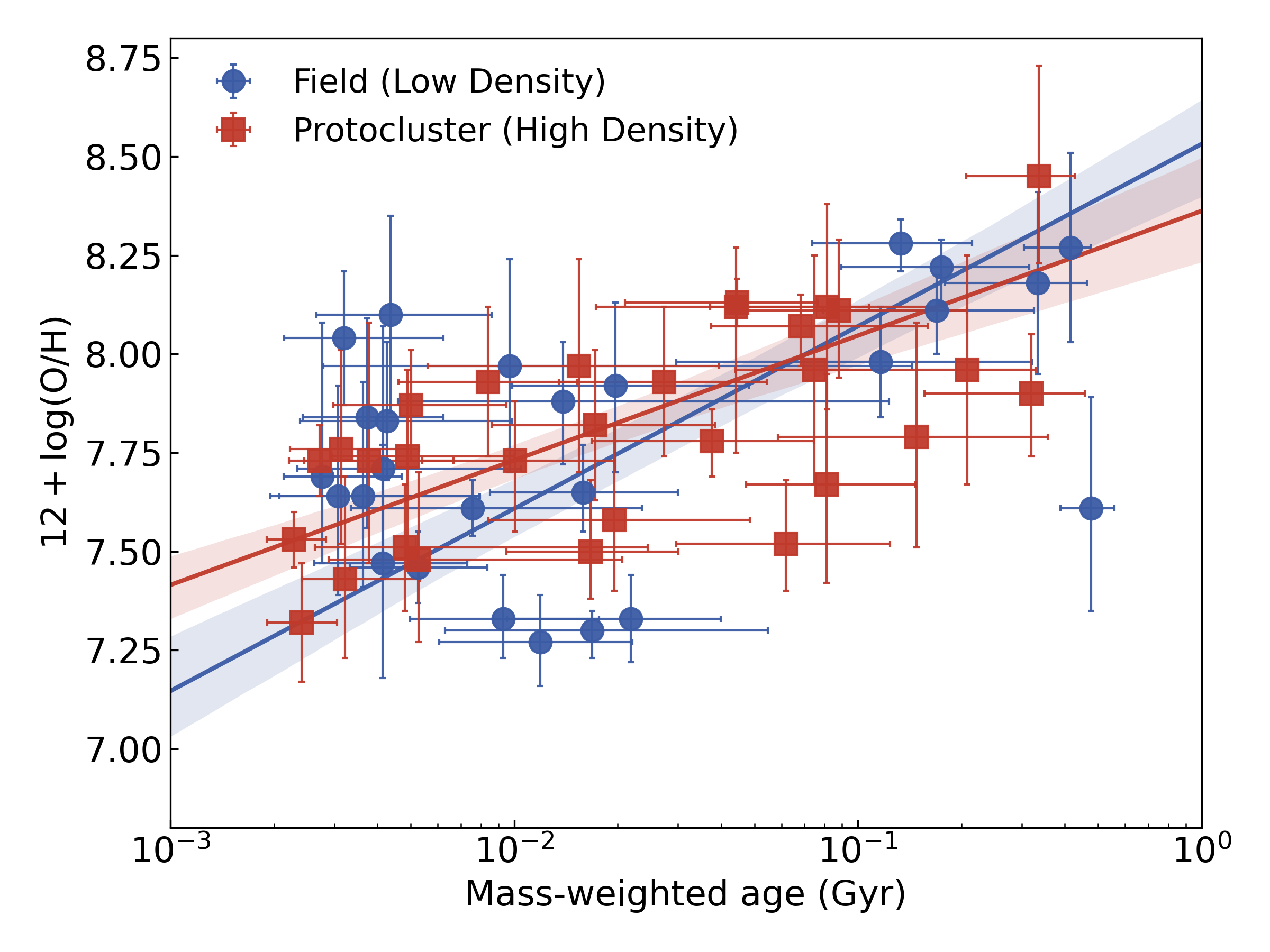}
    \caption{Age--metallicity relation for galaxies in low- and high-density environments in the CEERS and JADES fields. Blue circles represent field galaxies (low-density environments), while red squares denote protocluster galaxies (high-density environments). The solid lines show the best-fit relations obtained from orthogonal distance regression in the $\log_{10}(t_{\mathrm{mw}}/\mathrm{Gyr})$--metallicity plane, accounting for uncertainties in both variables. The shaded regions correspond to the $16$th--$84$th percentile confidence bands derived from 2000 bootstrap realizations. 
    Protocluster galaxies appear to show a flatter slope than field galaxies, although this may largely reflect increased scatter in the relation. This broader spread likely arises from more varied chemical enrichment histories in overdense regions at early cosmic times.}
    \label{fig:age_metallicity_env}
\end{figure}

Figure~\ref{fig:age_metallicity_env} presents the resulting age--metallicity relation for field and protocluster galaxies. Both subsamples exhibit a positive correlation between stellar population age and metallicity, indicating progressive chemical enrichment with increasing stellar age. It is consistent with expectations from chemical evolution models, in which metallicity increases as galaxies convert gas into stars and recycle metals through feedback and inflows \citep[e.g.,][]{Finlator2008,Dave2012,Lilly2013}. For field galaxies, we obtain a best-fit slope and intercept of $a_{\mathrm{pc}} = 0.46 \pm 0.08$ and $b_{\mathrm{pc}} = 8.53 \pm 0.14$, with bootstrap confidence intervals of $a_{\mathrm{pc}} = 0.46^{+0.07}_{-0.08}$ and $b_{\mathrm{pc}} = 8.53^{+0.11}_{-0.13}$. For protocluster galaxies, the corresponding values are $a_{\mathrm{field}} = 0.32 \pm 0.05$ and $b_{\mathrm{field}} = 8.36 \pm 0.10$, with bootstrap confidence intervals of $a_{\mathrm{field}} = 0.32^{+0.07}_{-0.07}$ and $b_{\mathrm{field}} = 8.37^{+0.13}_{-0.14}$. These results indicate that field galaxies follow a steeper age--metallicity relation. The slope difference is therefore $\Delta a = a_{\mathrm{pc}} - a_{\mathrm{field}} = 0.14$, with a combined uncertainty of $\sigma_{\Delta a} = \sqrt{\sigma_{\mathrm{pc}}^2 + \sigma_{\mathrm{field}}^2} \simeq 0.09$, corresponding to a significance level of $\sim 1.6\sigma$. Although moderate, this difference is consistently recovered by both the orthogonal distance regression and the bootstrap resampling analysis, indicating that the trend is not driven by statistical fluctuations or a small number of outliers.

The observed environmental trend also connects naturally to the mass--metallicity relation and its redshift evolution. Previous studies have reported environmental variations in the mass--metallicity relation at both low and intermediate redshift, with galaxies in dense environments exhibiting systematically higher metallicities at fixed stellar mass (e.g. \citealt{Peng2014,Darvish2015,Chartab2021}). 
Our results extend this picture to the epoch of reionization, demonstrating that environmental effects influence not only the instantaneous metallicity but also its evolution with galaxy age. In this sense, the age--metallicity relation can be interpreted as a ``chemical clock'' that encodes the assembly history of galaxies in different environments.

A steeper age--metallicity relation in the field can be interpreted as evidence for a more extended and gradual chemical enrichment history. In relatively isolated environments, galaxies are likely to experience prolonged gas accretion and episodic star formation, such that metallicity grows steadily as stellar populations age. In contrast, the shallower relation observed in protocluster galaxies suggests a more rapid early enrichment followed by a slower evolution at later times. Galaxies in overdense regions are expected to assemble earlier and experience enhanced early star formation, leading to elevated metallicities at young ages and a reduced sensitivity of metallicity to subsequent stellar aging. Such behaviour is qualitatively consistent with hierarchical structure formation models, in which protocluster regions collapse earlier and host galaxies with accelerated baryon cycling and enhanced metal retention \citep[e.g.,][]{Chiang2017,Kannan2023,Garcia2024}. In this scenario, metallicity becomes partially decoupled from stellar age once early enrichment has already taken place.

The environmental contrast seen in the age--metallicity plane therefore complements the trends observed in the MZR and FMR. While overdense galaxies tend to be more metal-rich at fixed stellar mass, their chemical evolution appears less strongly correlated with stellar age than in the field. This suggests that environmental effects primarily operate by accelerating the early phase of chemical enrichment rather than by sustaining a long-term enhancement in enrichment efficiency. Similar behaviour has been suggested in both observations and simulations, where galaxies in dense environments exhibit earlier assembly histories and reduced sensitivity to subsequent evolutionary processes \citep[e.g.,][]{Steidel2005,Overzier2016,Dave2020}.

However, several caveats must be considered when interpreting the age--metallicity relation at $z>4$. Mass-weighted ages derived from SED fitting are subject to degeneracies between age, metallicity, dust attenuation, and star formation history, which can systematically affect the inferred slopes \citep[e.g.,][]{carnall2019vandels,Leja2019}. 
In addition, the protocluster and field samples occupy partially different regions of stellar mass and SFR parameter space, which may introduce residual correlations between age and metallicity through their mutual dependence on mass. Finally, the limited sample size and substantial measurement uncertainties imply that the slope difference between environments is statistically modest, and should be interpreted as suggestive rather than definitive. Despite these limitations, the observed environmental dependence of the age--metallicity relation provides independent evidence that the timing of chemical enrichment differs between galaxies in dense and diffuse regions during the epoch of reionization. 

\section{Discussion}\label{sec:5}
\subsection{Comparison with radiation–hydrodynamics simulations}

%\begin{figure}
%\centering
%\includegraphics[width=\columnwidth]{plot/3d.pdf}
%\caption{3D spatial distribution of simulated galaxies in the \textsc{sphinx} cosmological simulation. Red and blue points represent galaxies residing in overdense (cluster) and underdense (field) environments, respectively, identified based on their local number density within a $10^3$ grid. }
%\label{fig:3d}
%\end{figure}

To interpret the observed environmental trends in the MZR within a physical framework, we compare our results with the \textsc{SPHINX} radiation--hydrodynamics simulations \citep{Rosdahl2018,Katz2022}. We use the publicly released \textsc{SPHINX} galaxy catalogue and define environment in three dimensions by discretising the simulation volume into a $10^3$ grid and assigning each galaxy a local number density based on its host cell. We then select galaxies in the lowest and highest quartiles of this density distribution as the \emph{underdense} and \emph{overdense} subsamples, respectively, thereby maximising environmental contrast. While this 3D definition is not identical to our projected $\Sigma_{5}$ used for the JWST samples, it provides a clean and physically motivated separation in large-scale density that is not affected by projection or redshift uncertainties.

\begin{figure*} %mz-SPHINX-only-v1.py
\centering
\includegraphics[width=0.95\textwidth]{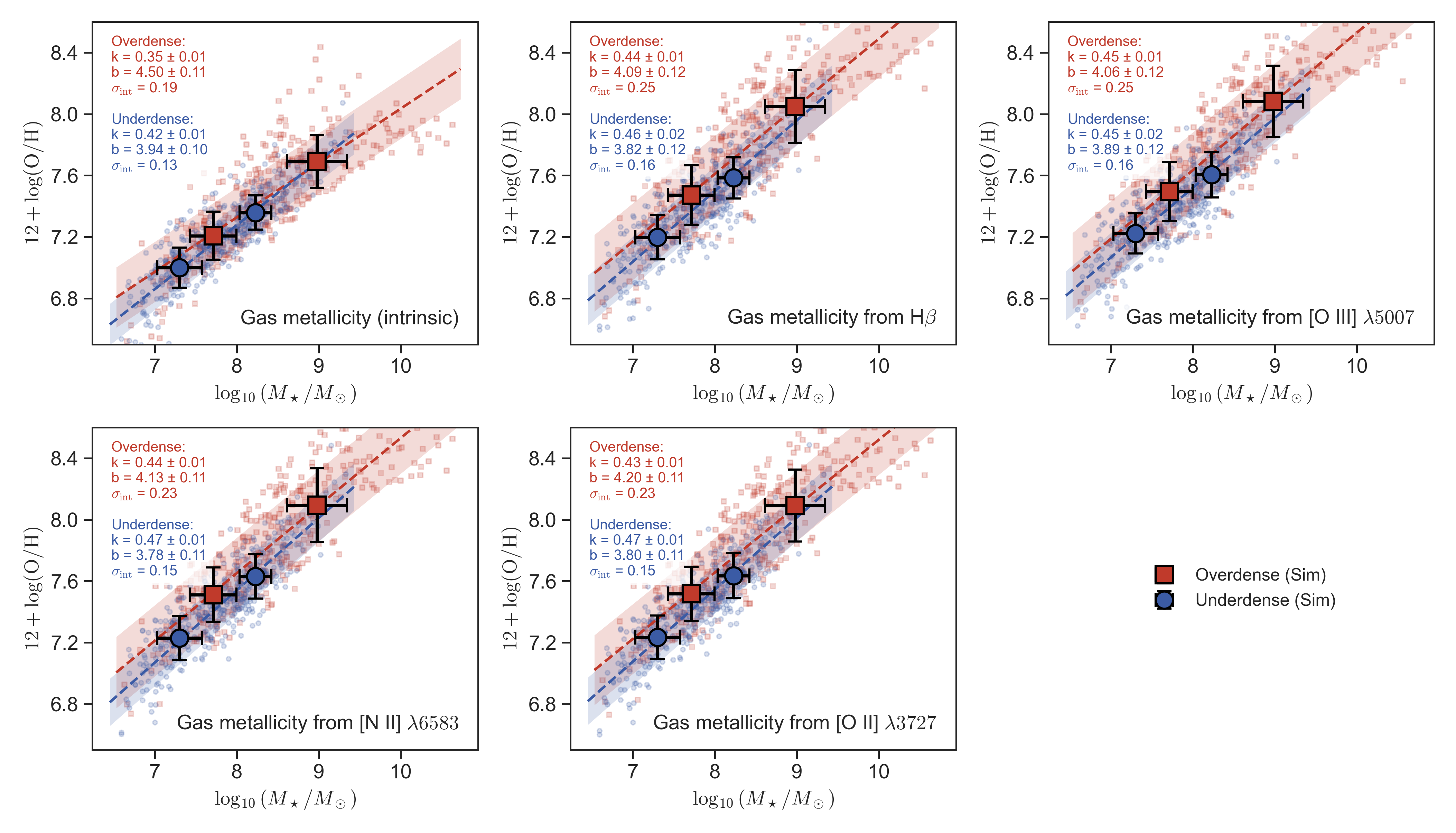}
\caption{Environmental dependence of the mass--metallicity relation  in the \textsc{SPHINX} simulations.
Panels show the stellar mass versus gas-phase metallicity, expressed as $12+\log(\mathrm{O/H})$, derived using different nebular-line-based metallicity indicators.
Galaxies in overdense and underdense environments are shown by red squares and blue circles, respectively.
Small symbols represent individual galaxies, while large symbols with error bars denote binned mean values, where the error bars indicate the uncertainty on the mean.
Dashed lines show best-fitting linear relations for each environment, and the shaded bands indicate the $1\sigma$ confidence intervals of the fits.
The metallicity indicator adopted in each panel is indicated in the bottom-right corner.}
\label{fig: SIM_SUM}
\end{figure*}

Figure~\ref{fig: SIM_SUM} shows the \textsc{SPHINX} MZR measured using five metallicity proxies available in the catalogue: the intrinsic gas metallicity, and four nebular-line-based estimates tied to H$\beta$, [O\,III]~$\lambda5007$, [N\,II]~$\lambda6583$, and [O\,II]~$\lambda3727$. For ease of comparison to observations, we express metallicity on the oxygen-abundance scale, $Z \equiv 12+\log(\mathrm{O/H})$, by applying a constant solar offset ($12+\log(\mathrm{O/H})_\odot=8.69$; \citealt{Asplund2009}) to the tabulated values in the catalogue. In each panel we fit a linear relation $Z=k\log(M_\star/M_\odot)+b$ to the overdense and underdense subsamples. The relations are fitted using weighted least squares, assuming a constant metallicity uncertainty of 
$\sigma_Z = 0.05~\mathrm{dex}$ for all galaxies.

First, we find that \textsc{SPHINX} predicts a systematic metallicity enhancement in overdense regions at fixed stellar mass for all four nebular-line-based proxies. Using the best-fitting relations, the overdense--underdense offset at $\log(M_\star/M_\odot)=8$ is $\Delta Z \simeq 0.11$--$0.12$\,dex for the H$\beta$, [O\,III], [N\,II], and [O\,II] estimators, and remains positive at $\log(M_\star/M_\odot)=9$ with $\Delta Z \simeq 0.07$--$0.10$\,dex. Second, the underdense subsample consistently shows a slightly steeper MZR slope ($a_{\rm under}\simeq0.42$--0.47) than the overdense subsample ($a_{\rm over}\simeq0.36$--0.45) for the line-based metallicities, implying that the environmental offset is larger at lower masses and decreases mildly toward higher masses. The fact that the sign and approximate amplitude of $\Delta Z$ persist across multiple line-based metallicity proxies indicates that the environmental signal is not an artefact of any single diagnostic, but reflects a genuine difference in enrichment histories between galaxies forming in distinct large-scale regions. The intrinsic gas metallicity panel shows a weaker (and at the high-mass end, sign-changing) offset, indicating that the environmental signal is clearest when metallicity is tied to nebular/line-emitting gas, i.e.\ the component most directly comparable to observational strong-line metallicities. 

\begin{figure} %mz-SPHINX-V2.py
\centering
\includegraphics[width=\linewidth]{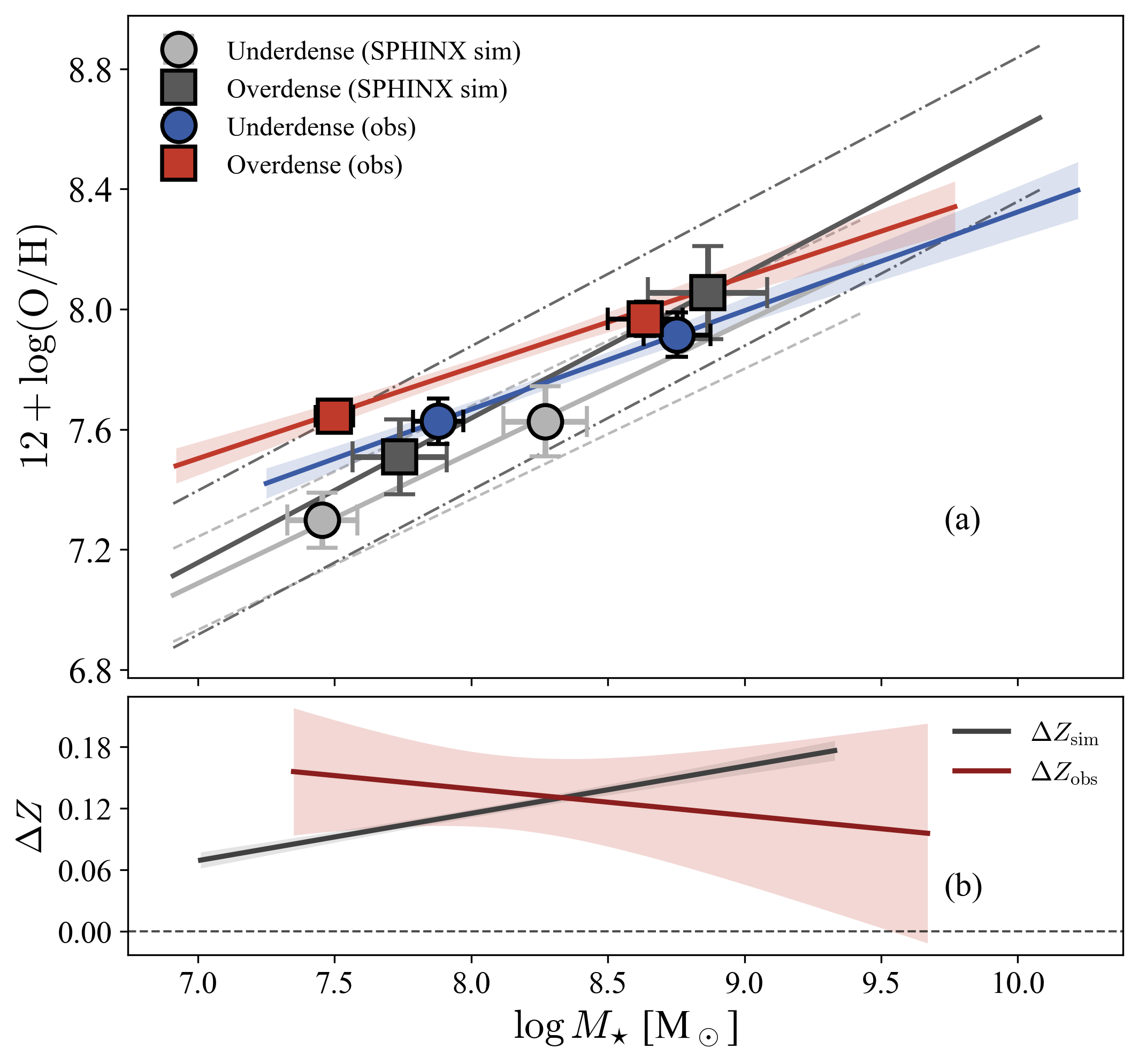}
\caption{
Mass--metallicity relation (MZR) comparison between SPHINX simulations and JWST observations 
(CEERS and JADES) at $z \sim 4$--7 in different environments.
Blue circles and red squares show the observational results for underdense (field) and overdense (protocluster) galaxies, respectively. Grey circles and squares show the corresponding 
SPHINX simulation results. Solid lines indicate the best-fitting linear relations for each sample. For the SPHINX simulations, dashed curves around the fit represent the prediction interval, computed from the parameter covariance of the fit combined with the intrinsic scatter estimated from the residuals. Shaded bands around the observational relations show the uncertainty from the fitting covariance. Large symbols denote the mean metallicity in two stellar-mass bins for each environment. For the simulations, the error bars represent the $1\sigma$ scatter within each bin, while 
for the observations they represent the uncertainty on the mean.
The lower panel shows the environmental metallicity offset, defined as $\Delta Z \equiv Z_{\rm cluster} - Z_{\rm field}$, for both simulations and observations. The curves are derived from the corresponding best-fitting relations in the upper panel, and the shaded regions indicate the propagated uncertainties from the fit covariances.}
\label{fig:mzr_obs_sim_env}
\end{figure}

In Figure~\ref{fig:mzr_obs_sim_env}, we compare the observed mass--metallicity relation (MZR) with predictions from the SPHINX cosmological simulations. These simulation predictions align well with our observational inference from CEERS+JADES at $4.5<z<7$ (Sec.~\ref{sec:3.1}). we find a mild metallicity enhancement in overdense environments at fixed mass (bootstrap--MCMC median $\Delta b \simeq 0.14$\,dex with $\Pr(\Delta b>0)=0.89$), and slopes that are broadly consistent within uncertainties. In this sense, \textsc{SPHINX} provides a consistent theoretical counterpart in which overdense galaxies reach higher chemical maturity earlier, while the mass dependence of the offset remains modest. The amplitude seen in \textsc{SPHINX} for the nebular proxies ($\Delta Z\sim0.1$\,dex at $\log M_\star\sim8$) is comparable to the scale of the environmental signal suggested by our JWST sample, supporting an interpretation in which accelerated enrichment and/or enhanced metal retention operates may already during the epoch of reionization.

Our results also connect to a broader body of work on environmental metallicity trends. At low redshift, galaxies in groups/clusters show modest metallicity enhancements at fixed mass ($\sim$0.05--0.1\,dex), often interpreted as reduced pristine inflow, enhanced recycling, or pre-processing \citep[e.g.][]{Peng2015,Darvish2015,Chartab2020}. At $z\sim2$--3, several protocluster studies report evidence for accelerated enrichment and/or reduced scatter relative to field populations, although sample sizes and selection effects remain limiting \citep[e.g.][]{Shimakawa2015,Chartab2021,Kashino2022}. The \textsc{SPHINX} trend at $z\gtrsim5$ is therefore qualitatively consistent with a picture in which dense regions assemble earlier and process gas more efficiently, imprinting a measurable chemical offset that persists across cosmic time. Earlier halo collapse and higher accretion/merger rates can raise the time-integrated star
formation and enhance metal production, while deeper potentials and more efficient recycling can increase metal retention
\citep[e.g.][]{Finlator2008,Lilly2013,Chiang2017,Torrey2019,Dekel2023}. 

We caution that a comparison between \textsc{SPHINX} and our data is limited by several factors. First, the observational environment metric is projected and evaluated in a finite redshift slice, whereas \textsc{SPHINX} uses a 3D density field; this can dilute the observed contrast relative to the intrinsic 3D separation. Second, strong-line metallicity estimates at $z>4$ can be sensitive to evolving ionization conditions and abundance patterns \citep[e.g.][]{Nakajima2014,Sanders2023,Nakajima2023,Curti2024}, and the simulation line-based metallicity proxies are not guaranteed to match the exact calibrations adopted for our JWST measurements.  For these reasons, the comparison is most robust when focusing on relative environmental trends. Both the simulations and our observations indicate that galaxies in overdense regions are systematically more metal enriched than field galaxies at fixed stellar mass in the early Universe.

\subsection{\texorpdfstring{Environmental Dependence of the Cosmic Star Formation Rate Density at \ensuremath{z\sim6\text{--}9}}{Environmental Dependence of the Cosmic Star Formation Rate Density at z~6--9}}

\label{sec:sfrd_environment}\label{sec:5.2}

\begin{figure}
    \centering
    \includegraphics[width=\columnwidth]{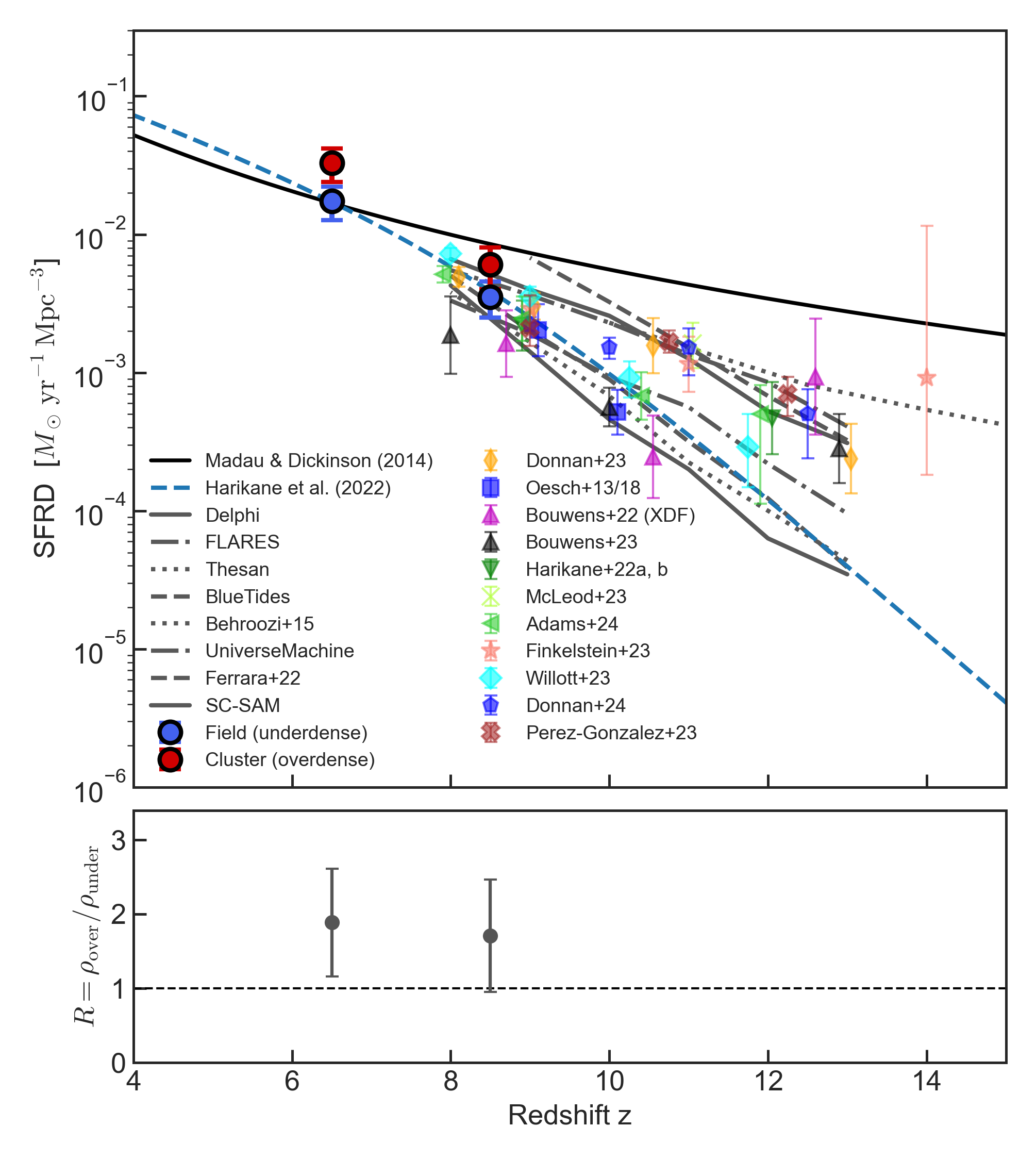}
    \caption{Environmental dependence of the cosmic star-formation rate density at $z\simeq6$--9. \emph{Top panel:} Cosmic star-formation rate density (SFRD) measured separately in underdense (field; blue) and overdense (cluster; red) environments using CEERS and JADES galaxies. The SFRD is computed with a $V_{\max}$ formalism, error bars and shaded regions indicate $1\sigma$ uncertainties estimated via bootstrap resampling. Solid and dashed curves show literature measurements of the global SFRD from \citet{MadauDickinson2014} and \citet{Harikane2022}, respectively. Additional literature data points and symbols are described in Section~\ref{sec:5.2}. \emph{Bottom panel:} Ratio of the SFRD in overdense to underdense environments, $R=\rho_{\rm over}/\rho_{\rm under}$, as a function of redshift. The dashed horizontal line marks $R=1$, corresponding to no environmental difference. }
    \label{fig:sfr_density}
\end{figure}

In this section we quantify how large-scale environment regulates early galaxy formation. This requires 
measuring the cosmic star formation rate density (SFRD) separately in overdense and 
underdense regions. Using the CEERS and JADES photometric-selected galaxy samples 
(see Sec.~\ref{sec:data}), we compute the SFRD in both environments following a 
$V_{\max}$ formalism that incorporates survey completeness and luminosity-dependent 
detectability limits. The resulting SFRD measurements for field (underdense) and 
cluster (overdense) environments are presented in Fig.~\ref{fig:sfr_density}.  
These measurements represent one of the first direct comparison of the SFRD between 
environmentally distinct regions during the epoch of reionization.

For each galaxy, we derive the accessible comoving volume $V_{\max}$ based on the 
maximum redshift $z_{\max}$ at which it would satisfy the survey's limiting magnitude, 
computed from its rest-frame UV absolute magnitude and the appropriate distance modulus. 
Survey completeness is incorporated through a weight 
$w=1/\mathrm{max}(c_{\rm pl},0.05)$, where $c_{\rm pl}$ is the catalog completeness 
estimated from the injection--recovery simulations provided by \citet{Adams2024} and \citet{Conselice2025}. 
Galaxies are classified as overdense or underdense via the percentile-based 
$\Sigma_{5}$ thresholding described in Sec.~\ref{sec:environment}.  
The SFRD in a redshift interval $[z_{1},z_{2}]$ is then
\begin{equation}
\rho_{\rm SFR} = 
\sum_{i}\frac{w_{i}\,\mathrm{SFR}_{i}}{V_{\max,i}}.
\end{equation}
Uncertainties are estimated through non-parametric bootstrap resampling 
($N=50$--100 realizations), ensuring robustness against outliers and sample variance.

%\subsubsection{Comparison with Literature SFRD Constraints}

Figure~\ref{fig:sfr_density} compares our environmental star formation rate densities (SFRDs) with several other determinations of the 
global cosmic SFRD.  
We overlay the analytic parameterization of \citet{MadauDickinson2014},  
the updated high-redshift constraints from \citet{Finkelstein2016,Oesch2018},  
the recent JWST-based luminosity function results from \citet{Bouwens2022,Bouwens2023,Adams2024,donnan2023evolution,Donnan2024,Harikane2022,Harikane2022b,McLeod2023,finkelstein2023ceers,PerezGonzalez2023},  
and the empirical fit from \citet{Harikane2022} %, whose functional form is
%
%\begin{equation}
%\rho_{\rm SFR}(z) = 
%\frac{1}{61.7(1+z)^{-3.13}
%+ 10^{0.22(1+z)}
%+ 2.4\times 10^{0.50(1+z)-3.0}}
%\label{eq:harikane22_sfrd}
%\end{equation}
%
%and provides an excellent description 
 of the global SFRD evolution at 
$4 \lesssim z \lesssim 12$.

Our measurements lie broadly within the envelope spanned by the above literature 
compilations.  
However, a clear environmental separation emerges: in all redshift bins we measure 
a systematically elevated SFRD in overdense regions compared to underdense regions 
(Figure~\ref{fig:sfr_density}).  
We quantify this enhancement through the ratio
\begin{equation}
R(z) \equiv 
\left( \frac{\rho_{\rm over}(z)}{\rho_{\rm under}(z)} \right).
\end{equation}
Across $z\simeq6$--9, we find $R\approx2$, with bootstrap uncertainties 
preserved using the same resampling procedure.  
This result implies that the most actively star-forming galaxies at these epochs 
preferentially reside in the densest early structures, consistent with models in 
which early-forming halos experience accelerated gas accretion and more rapid 
assembly in protocluster environments. Given the limited survey volumes of CEERS and JADES, cosmic variance is expected to contribute significantly to the uncertainties. 
Nevertheless, the overdense-to-underdense contrast remains consistently above unity across all redshift bins, suggesting that the trend is unlikely to be driven solely by statistical fluctuations.

The enhanced SFRD observed in overdense regions is broadly consistent with theoretical expectations for the early assembly of large-scale structure. In the framework of biased galaxy formation, overdense regions correspond to the high peaks of the primordial density field, where dark-matter halos collapse earlier and grow more rapidly than in average-density environments \citep[e.g.,][]{Chiang2017,Overzier2016}. These halos are therefore expected to experience higher mass accretion rates and elevated star-formation duty cycles, naturally leading to an enhanced contribution to the cosmic star-formation budget.

Hydrodynamical simulations further predict that protocluster environments are characterised by frequent mergers, sustained cold gas inflows along filaments, and efficient recycling of baryons, all of which can amplify star formation at early epochs \citep[e.g.,][]{Kannan2022,Wilkins2023,Behroozi2013}. In this context, the elevated $\rho_{\rm SFR}$ measured in overdense regions can be interpreted as the combined outcome of accelerated halo growth and intensified baryon cycling, rather than solely an increase in star-formation efficiency at the level of individual galaxies.

The concentration of star formation in dense regions has important implications for cosmic reionization. If a substantial fraction of early star formation is spatially clustered within protocluster environments, the resulting ionizing photon production is expected to be highly inhomogeneous, potentially giving rise to large ionized bubbles and a patchy reionization topology \citep[e.g.,][]{Furlanetto2004,Madau2015,qiong2026}. 

Taken together, the measurements in Fig.~\ref{fig:sfr_density} support a picture in which environment influences galaxy growth already within the first billion years after the Big Bang. Rather than merely tracing regions of higher galaxy density, overdense environments appear to host accelerated assembly histories and enhanced baryon processing, leading to a significant contribution to the star-formation activity.

\section{Conclusions}\label{sec:6}

We have used deep \textit{JWST} observations from the CEERS and JADES fields to investigate how large-scale environment influences the chemical enrichment and star-formation activity of galaxies at $z\sim4$--10. By combining robust measurements of stellar mass, star-formation rate, gas-phase metallicity, galaxy size, and local density, we provide a unified view of the environmental imprint on early galaxy evolution. Our main conclusions are as follows.

\begin{enumerate}
\item Using a percentile-based definition of the projected fifth-nearest-neighbour surface density, we find that galaxies in overdense regions exhibit a modest but systematic metallicity enhancement relative to field galaxies at fixed stellar mass. A bootstrap--MCMC analysis of the mass--metallicity relation yields an intercept offset of $\sim0.1$--0.2\,dex in favour of overdense environments, with high posterior probability, while the slopes of the relation remain consistent within uncertainties. This indicates accelerated chemical enrichment in the densest regions already at $z\gtrsim5$.

\item The environmental separation becomes more pronounced when star formation is included. Both the $\mu$-based formulation and the fitted fundamental metallicity plane reveal a stronger dependence of metallicity on star-formation rate in overdense regions, together with a smaller intrinsic scatter. These results suggest that galaxies in dense environments experience more tightly regulated gas processing and more efficient metal retention than their field counterparts.

\item After removing the primary dependence on stellar mass, metallicity residuals show no statistically significant correlation with instantaneous star-formation rate or local density. This implies that the observed environmental trends arise predominantly from differences in the global growth histories of galaxies rather than from short-timescale variations in star formation.

\item We find a weak metallicity–size relation, with metallicity increasing slightly with size and flattening at larger radii. At fixed stellar mass, this dependence is modest and shows little environmental variation, indicating that size plays a secondary role in chemical enrichment.

\item We find a positive age–metallicity relation at $5<z<10$, indicating ongoing chemical enrichment with time. The relation is steeper in the field, suggesting more extended enrichment histories compared to overdense regions.

\item Using a $V_{\max}$ approach with completeness corrections, we find that the cosmic star-formation rate density at $z\simeq6$--9 is enhanced in overdense regions by a factor of $\sim2$ compared to underdense regions. This indicates that a disproportionate fraction of early star formation occurred in the progenitors of present-day clusters, with important implications for the topology and timing of cosmic reionization.
\end{enumerate}

Taken together, our results support a scenario in which environmental effects on galaxy evolution are already established during the epoch of reionization. Galaxies in overdense regions assemble earlier, form stars more efficiently, and enrich their interstellar medium more rapidly than field systems, leaving detectable imprints on both the mass--metallicity and star-formation relations. While current samples remain limited by cosmic variance and metallicity calibration uncertainties, ongoing and upcoming \textit{JWST} spectroscopy over wider areas, combined with ALMA and future facilities, will enable decisive tests of environment-driven chemical evolution in the early Universe.

\section*{Acknowledgements}

We acknowledge support from the ERC Advanced Investigator Grant EPOCHS (788113) and support from STFC studentships. 
This work is based on observations made with the NASA/ESA \textit{Hubble Space Telescope} (HST) and NASA/ESA/CSA \textit{James Webb Space Telescope} (JWST) obtained from the \texttt{Mikulski Archive for Space Telescopes} (\texttt{MAST}) at the \textit{Space Telescope Science Institute} (STScI), which is operated by the Association of Universities for Research in Astronomy, Inc., under NASA contract NAS 5-03127 for JWST, and NAS 5–26555 for HST. The observations used in this work are associated with JWST program 1345, 1180 1176, and 2738. The authors thank all involved in the construction and operations of the telescope as well as those who designed and executed these observations.

This work makes use of {\tt astropy} \citep{Astropy2013,Astropy2018,Astropy2022}, {\tt matplotlib} \citep{Hunter2007}, {\tt reproject}, {\tt DrizzlePac} \citep{Hoffmann2021}, {\tt SciPy} \citep{2020SciPy-NMeth}, {\tt photutils} \citep{Bradley2022}, and {\tt galfind} (\doi{10.5281/zenodo.18613231}).

%%%%%%%%%%%%%%%%%%%%%%%%%%%%%%%%%%%%%%%%%%%%%%%%%%
\section*{Data Availability}

The JWST data used in this work are available in the Cosmic Evolution Early Release Science Survey (ID: 1345, PI: S. Finkelstein, \citealt{Bagley2023}), the JWST Advanced Deep Extragalactic Survey ( ID: 1180 and 1210, PI: Eisenstein, N. Lützgendorf, \citealt{Rieke2023} ), through the Mikulski Archive for Space Telescopes (https://mast.stsci.edu/).  Additional data products will be shared on reasonable request to the first author.

%%%%%%%%%%%%%%%%%%%% REFERENCES %%%%%%%%%%%%%%%%%%

% The best way to enter references is to use BibTeX:

\bibliographystyle{mnras}
\bibliography{example} % if your bibtex file is called example.bib

% Alternatively you could enter them by hand, like this:
% This method is tedious and prone to error if you have lots of references
%\begin{thebibliography}{99}
%\bibitem[\protect\citeauthoryear{Author}{2012}]{Author2012}
%Author A.~N., 2013, Journal of Improbable Astronomy, 1, 1
%\bibitem[\protect\citeauthoryear{Others}{2013}]{Others2013}
%Others S., 2012, Journal of Interesting Stuff, 17, 198
%\end{thebibliography}

%%%%%%%%%%%%%%%%%%%%%%%%%%%%%%%%%%%%%%%%%%%%%%%%%%

%%%%%%%%%%%%%%%%% APPENDICES %%%%%%%%%%%%%%%%%%%%%

%\appendix

%\section{Some extra material}

%If you want to present additional material which would interrupt the flow of the main paper, it can be placed in an Appendix which appears after the list of references.

%%%%%%%%%%%%%%%%%%%%%%%%%%%%%%%%%%%%%%%%%%%%%%%%%%

% Don't change these lines
\bsp	% typesetting comment
\label{lastpage}
\end{document}